\documentclass[twoside]{article}
\usepackage{amssymb,amsmath,multirow,epsfig,graphicx,color}

\usepackage{epsfig}
\usepackage{graphicx,amsmath}
\def\be{\begin{eqnarray} &&}

\def\ee{\end{eqnarray}}

\usepackage{subfigure}
\usepackage{booktabs}
\usepackage{topcapt}

\def\sla#1{\rlap\slash #1}

\usepackage{soul}
\usepackage{graphicx}
\usepackage{dcolumn}
\usepackage{bm}
\usepackage{ulem}

\usepackage{booktabs}
\usepackage{topcapt}

\newcommand{\feynslash}[1]{/\hspace*{-2mm} #1}

\begin{document}
\thispagestyle{plain}
\begin{center}
{\Large \bf \strut
The soft-gluon limit and the infrared enhancement of the quark-gluon vertex 
\strut}\\
\vspace{10mm}
{\large \bf 
  Orlando Oliveira$^a$
  Tobias Frederico$^b$,
  Wayne de Paula$^b$,
}
\end{center}

\noindent{
{\small $^a$\it  CFisUC, Departamento de F\'{\i}sica, Universidade de Coimbra, 3004-516 Coimbra, Portugal} \\
\small $^b$\it Instituto Tecnol\'ogico da Aeron\'autica, DCTA, 12.228-900 S\~ao Jos\'e dos Campos,  Brazil} \\

\date{Received: date / Revised version: date}
%

\begin{abstract}
The Schwinger-Dyson quark equation (SDE) combined with results from lattice simulation for the propagators
are used to obtain information on the quark-gluon vertex,
taking into account the recent full QCD lattice results for the soft-gluon limit. 
Its inclusion leads to a clear enhancement of the infrared quark-gluon vertex.
We also find that the relative contribution of the quark-ghost kernel to the quark-gluon vertex 
in the infrared region does not follow the rules from the perturbative analysis of the ultraviolet region. 
This shows that for QCD the intuition based on perturbation theory does not apply to the full momentum range.
The framework developed in the current work provides analytical expressions for all the longitudinal components of vertex taken into account.
\end{abstract}


\section{ Introduction and Motivation} 

The interaction between quarks and gluons is described by Quantum Chromodynamics (QCD). 
If at high energies asymptotic freedom allows the use of perturbation theory, at low energies QCD becomes strongly coupled and 
other type of techniques are needed to solve the theory.
Confinement and chiral symmetry breaking are two important non-perturbative outcomes of QCD  associated with its infrared (IR) 
properties \cite{Alkofer:2000wg,Fischer:2006ub}.
In particular,
confinement impacts on the computation of hadronic transitions elements also at high energies.
 
 Lattice QCD simulations is a successful \textit{ab initio}  
approach to study the non-perturbative regime.
The use of continuous methods, such as Schwinger-Dyson equations (SDE), Bethe-Salpeter, Faddeev equations,
functional re\-nor\-ma\-li\-sa\-tion group techniques and Holographic models complement our catalogue of non-perturbative tools for
quantum field theories.

The gluon, the ghost and the quark propagators have been thoroughly studied within the non-perturbative re\-gi\-me of QCD
and, nowadays, one has a fair understanding and description of these two point correlation functions. 
On the other hand, our level of knowledge on the three particle vertices is much poor when compared to the two point functions.
 Among the QCD three particle vertices is the quark-gluon vertex that plays a major role in our understanding
of hadrons.

Herein we pursuit the efforts initiated in \cite{Rojas:2013tza} and developed further in \cite{Oliveira:2018ukh,Oliveira:2018fkj} that
combine results from lattice simulation in the Landau gauge, for the propagators, and the quark gap equation, 
together with an exact solution of the Slavnov-Taylor identity for the vertex, to get information on the 
quark-gluon vertex. The exact solution of the Slavnov-Taylor identity for the vertex calls for the quark-ghost kernel,
a four point Green function, that is described in terms of four form factors~\cite{Davydychev:2000rt}, our unknowns.
The main novelty in the present work is the inclusion of the lattice results for the soft-gluon limit of the quark-gluon vertex,
that translate into normalisation conditions for the quark-ghost kernel form factors. In practice, this was achieved 
by building an ansatz that is inspired on the analysis of the lattice data for the quark-gluon vertex \cite{Oliveira:2018ukh} and
the gap equation is solved for the form factors, written as  Pad\'e approximants. 
We stress that all other quantities in the gap equation are known from lattice simulations, including the quark propagator in the Landau gauge.
The resulting quark-gluon form factors are identical to those computed previously \cite{Oliveira:2018fkj} with the exception of the infrared region,
where it is observed a clear enhancement.
Furthermore,  
a detailed analysis of the relative contribution  of the quark-ghost kernel form factors to the quark-gluon vertex  
does not comply, in the infrared region, with their ultraviolet (perturbative) behaviour.  

This paper reports on results for the quark-gluon vertex as outlined above and it is organized as follows. 
In section 2, we set our notation for the quark gap equation and the quark-gluon vertex, where as usual we consider only the
longitudinal components, written in terms of the quark-ghost kernel form factors, which are parametrized by Pad\'e approximants
constrained by the
soft-gluon limit results from lattice simulations. 
In section 3, the Schwin\-ger-Dyson equation is decomposed for the vector and scalar parts of the quark self energies,
and the minimal requirements for the ultraviolet behaviour of the quark-ghost kernel are discussed  together with 
its anzats. Also, the explicit form of our anzats\"e for the longitudinal form factors are provided.  In
section 4, we  built our anzats such that the soft-gluon limit from QCD lattice simulations of $\lambda_1$ is incorporated. In section 5,
we present the results from  the inversion of the Schwinger-Dyson equations to get the coefficients of the Pad\'e approximants for the quark-ghost kernel 
 relying on simulating annealing to minimize the sum of the 
relative error of the scalar and vector equations.
In  section 6 we present our results for the form factors of the longitudinal components of the quark-gluon vertex and also analyze their
contribution of the  quark-ghost kernel separately. In section 7 we provide a discussion of our results to put in 
the perspective of some previous studies. 
This section is closed with  a summary of our work.

\section{The Quark Gap Equation and the Quark-Gluon Vertex}

The quark propagator is color diagonal and its spin-Lorentz structure reads, in Minkowski space,
\begin{eqnarray}
  S^{-1} (p) & = & -i \, \left( A(p^2) \feynslash  p\ -  B(p^2) \right) \nonumber \\
 & =  &  -i \, Z_2 (\, \feynslash  p - m^{\mathrm{bm}}) + \Sigma (p)
  \label{eq:SDE}
\end{eqnarray}
where $Z(p^2) = 1/A(p^2)$ is the quark wave function, 
$M(p^2) = B(p^2) / A(p^2)$ the renormalisation group invariant running
quark mass, $Z_2$ is the quark renormalisation constant and $m^{\mathrm{bm}}$ the bare current quark mass. The quark self-energy 
is given by
\begin{eqnarray}
 \Sigma (p)  &=&  Z_1 \, \int {d^4 q\over(2\pi)^4} ~ D_{\mu\nu}^{a b}(q)  ~ (\, i\, g\, t^b \gamma_\nu\,) ~ \nonumber\\
                      && \qquad\quad \times S(p-q) ~  \Gamma^a_\mu (-p, \, p-q, \, q) ,
 \label{DSEselfenergy}
\end{eqnarray}
where $Z_1$ is  a combination of several renormalisation constants and the Landau gauge gluon propagator is 
\begin{equation}
  D^{ab}_{\mu\nu} (q) =  -i\,  \delta^{ab} \left( g_{\mu\nu} - \frac{q_\mu q_\nu}{q^2} \right) D ( q^2 ) \ .
\end{equation}
The quark-gluon vertex is defined with incoming momenta  $p_1+p_2+p_3 = 0$, where $p_2$ is the incoming quark momentum, $-p_1$
the outgoing quark momentum and $p_3$ the incoming gluon momentum. Our notation follows that used in
\cite{Oliveira:2018ukh,Davydychev:2000rt}. 
The one-particle irreducible quark-gluon Green function is depicted as
\begin{equation}
 \Gamma^a_\mu (p_1, p_2, p_3) = g \, t^a \, \Gamma_\mu (p_1, p_2, p_3) \ , 
 \label{EQ:QGVertexLorentz}
\end{equation} 
where $g$ is the strong coupling constant and $t^a$ are the generators of the color SU(3) group in the fundamental representation. 

Assuming that the gluon propagator and $\Gamma^a_\mu$ are known, from the gap equation (\ref{DSEselfenergy}), one can get the quark propagator. 
If $Z(p^2)$ and $M(p^2)$ are known, it is possible to use (\ref{DSEselfenergy}) 
to extract information on the quark-gluon vertex. 
From the mathematical point of view, computing $\Gamma^a_\mu$ from the gap equation means solving  an ill-defined problem.
The introduction of a prior, that can be accommodated by regularising the integral equation or introducing a basis of functions, allows
to exactly and unambiguously solve the modified equation for the vertex. The solution depends on the prior and one should check 
its (in)dependence on the prior.

The vertex function $\Gamma_\mu$, see Eq. (\ref{EQ:QGVertexLorentz}), can be decomposed in a longitudinal 
$\Gamma^{(L)}_\mu$ and a transverse $\Gamma^{(T)}_\mu$ component, relative to the  gluon momenta, as
\begin{equation}
   \Gamma_{\mu}(p_1, \, p_2, \, p_3)  = \Gamma^{(L)}_{\mu}(p_1, \, p_2, \, p_3) + \Gamma^{(T)}_{\mu}(p_1, \, p_2, \, p_3)
\end{equation}   
and, by definition,  
\begin{equation}
 p_{3}^{\mu} ~ \Gamma^{(T)}_{\mu}(p_1, \, p_2, \, p_3)= 0 \ .
\end{equation} 
As is usual in the analysis of the Dyson-Schwinger equations, 
in the current work we will focus on the longitudinal component of the $\Gamma_\mu$ and will ignore $\Gamma^{(T)}_{\mu}$.
If a tensor basis for $\Gamma^{(L)}_\mu$ and $\Gamma^{(T)}_\mu$ is given, then $\Gamma_\mu$ is a sum 
of scalar form factors that multiply each of the elements of the tensor basis. The full vertex requires twelve form factors, with four of them
being  associated with the longitudinal component and eight with the transverse one. In the Ball-Chiu (BC) basis \cite{Ball:1980ay},
the longitudinal vertex is written as:
\begin{eqnarray}
   \Gamma^\mathrm{L}_\mu (p_1, p_2, p_3) & = & -i \, \bigg( \lambda_1  \, \gamma_\mu 
   ~ + ~ \lambda_2  \, ( \feynslash p_1 - \feynslash p_2 ) \left( p_1 - p_2 \right)_\mu \nonumber \\
   & & 
   + ~ \lambda_3 \, ( p_1 - p_2  )_\mu 
    ~ + ~ \lambda_4 \,  \sigma_{\mu\nu} \left( p_1 - p_2 \right)^\nu \bigg) , \nonumber\\
    \, 
   \label{EQ:Lvertex} 
\end{eqnarray}
where $\sigma_{\mu\nu} = {1\over 2} [\gamma_{\mu},\gamma_{\nu}]$ and $\lambda_i = \lambda_i(p^2_1, p^2_2,p^2_3)$. 
The transverse component of the BC vertex is written as 
\begin{eqnarray}
   \Gamma^\mathrm{T}_\mu (p_1, p_2, p_3)  & = & -i\, \sum^8_{i=1} \tau_i (p_1, p_2, p_3) ~ T^{(i)}_\mu (p_1 , p_2) \ ,
   \label{EQ:Tvertex} 
\end{eqnarray}
where the operators associated to the transverse part of the vertex are
\begin{eqnarray}\label{EQ:TFF}
 T^{(1)}_\mu (p_1 , p_2)  & = &   \left [ p_{1\mu} \left(p_2 \cdot p_3 \right) -  p_{2\mu} \left(p_1 \cdot p_3 \right) \right ] \mathbb{I}_D  \ , \nonumber \\
 T^{(2)}_\mu (p_1 , p_2)  & =  &   - T^{(1)}_\mu (p_1 , p_2)~ \left( \feynslash p_1 - \feynslash p_2 \right) \ ,  \nonumber \\
 T^{(3)}_\mu (p_1 , p_2)  & = &  p^2_3 \,  \gamma_\mu - p_{3\mu} \  \feynslash p_3 \ , \nonumber\\
 T^{(4)}_\mu (p_1 , p_2)  & = &     T^{(1)}_\mu  (p_1 , p_2) ~ \sigma_{\alpha\beta}\, p^\alpha_1 p^\beta_2 \ , \nonumber \\
 T^{(5)}_\mu (p_1 , p_2)  & = &   \sigma_{\mu\nu} \, p^\nu_3 \ ,  \nonumber \\
 T^{(6)}_\mu (p_1 , p_2)  & = &   \gamma_\mu \left( p^2_1 - p^2_2 \right) + \left( p_1 - p_2 \right)_\mu \,  \feynslash p_3 \ , \nonumber \\
 T^{(7)}_\mu (p_1 , p_2) & =  & -{1\over 2} \left( p^2_1 - p^2_2 \right) 
                      \left[ \gamma_\mu \left( \feynslash p_1 - \feynslash p_2 \right) - \left( p_1 - p_2 \right)_\mu \mathbb{I}_D \right]  \nonumber \\
                      & &
                      \hspace{1cm} -
                      \left( p_1 - p_2 \right)_\mu \sigma_{\alpha\beta} \, p^\alpha_1 \, p^\beta_2 \ , \hspace*{1cm} \nonumber \\
 T^{(8)}_\mu (p_1 , p_2) & =  &  - \gamma_\mu \, \sigma_{\alpha\beta} \, p^\alpha_1 \, p^\beta_2 + ( p_{1\mu} \, \feynslash  p_2 -  p_{2\mu} \, \feynslash p_1) \ .
\end{eqnarray}

The discrete symmetries of QCD constraint the quark-gluon vertex. For the longitudinal form factors,
charge conjugation invariance \cite{Davydychev:2000rt} requires
\begin{eqnarray}
   \lambda_i \left(p^2_1, \, p^2_2, \, p^2_3\right)  &=&   \lambda_i \left(p^2_2, \, p^2_1, \, p^2_3\right)|_{i=1,2,3}, \nonumber \\
   \lambda_4 \left(p^2_1, \, p^2_2, \, p^2_3\right)  &=&  -  \lambda_4 \left(p^2_2, \, p^2_1, \, p^2_3\right) \ .
 \label{Eq:VertexDiscretas}
\end{eqnarray}
These properties under interchange of quark momenta im\-ply that when $p^2_1 = p^2_2$  
the form factor
$\lambda_4 = 0$ as happens for the soft-gluon limit where $p_3 = 0$.

The  Slavnov-Taylor identity (STI) for the quark-gluon vertex, 
\begin{eqnarray}
 p^\mu_3 ~ \Gamma_\mu (p_1, \, p_2, \, p_3 ) = &&
F(p^2_3) \Big[ S^{-1}(-p_1) \,  H(p_1, \, p_2, \, p_3)\nonumber\\ && - \overline H(p_2, \, p_1, \, p_3) \,  S^{-1}(p_2)  \Big] ,
 \label{EQ:STI}
\end{eqnarray}
relates the longitudinal vertex form factors, $\lambda_i$'s $(i=1-4)$, with the quark propagator,  the quark-ghost kernels that are define by $H$ and $\overline H$, see~\cite{Davydychev:2000rt} 
for notation and definitions, and the ghost dressing function $F(p^2_3)$, related to the ghost propagator by
\begin{equation}
  D^{ab}(p^2) = - i \, \delta^{ab} \, F(p^2) / p^2 \ .
\end{equation}  
The quark-ghost kernel can be written in terms of four form factors \cite{Davydychev:2000rt}, 
called $X_i\equiv X_i(p^2_1, p^2_2, p^2_3)$ and $\overline X_i \equiv X_i (p^2_2, \, p^2_1, \, p^2_3 )$ , as
\begin{eqnarray}
   H(p_1, \, p_2, \, p_3 ) &=&  X_0   \mathbb{I}_D  + X_1   \, / \!\!\!p_1   + X_2   \, / \!\!\!p_2 
                  + X_3   \,\sigma_{\alpha\beta} \, p^\alpha_1 \, p^\beta_2  \ , \nonumber \\
  \overline H(p_2, \, p_1, \, p_3 ) & = &   X_0  \mathbb{I}_D     -   X_2   \, / \!\!\!p_1 
                  -    X_1  \, / \!\!\!p_2 
                  +  X_3  \,\sigma_{\alpha\beta} \, p^\alpha_1 \, p^\beta_2 \ .\nonumber\\
\end{eqnarray}
The STI (\ref{EQ:STI}) can be solved exactly for the  $\lambda_i$'s \cite{Aguilar:2010cn}  resulting in
\begin{small}
\begin{eqnarray}
 \lambda_1 & =  & \frac{F(p^2_3)}{2} \Bigg\{  
   A(p^2_1) \left[ X_0 + \left( p^2_1 - p_1 \cdot p_2 \right) X_3 \right] \nonumber \\
              && \qquad \quad
              +  B(p^2_1) \left[X_1 + X_2 \right] \Bigg\} + (p_1\leftrightarrow  p_2) \ ,
  \nonumber \\
 \lambda_2  &  = &
               \frac{F(p^2_3)}{2 (p^2_2 - p^2_1)}        \Bigg\{
               A(p^2_1)\! \left[ \left( p^2_1 + p_1\! \cdot p_2 \right) X_3  - X_0 \right] 
               \nonumber \\ 
               && \qquad \qquad \quad+  B(p^2_1) \left[X_2 - X_1 \right]   \Bigg\}+ (p_1\leftrightarrow  p_2) \ ,
 \nonumber \\
 \lambda_3   & =  &
                   \frac{F(p^2_3)}{p^2_1 - p^2_2}    \Bigg\{ 
                    A(p^2_1) \left[ p^2_1 \, X_1 + p_1 \cdot p_2 \ X_2\right]  +    B(p^2_1) \, X_0 
                   \Bigg\} \nonumber \\
           \nonumber      \\
 \lambda_4 &  = &  - \frac{F(p^2_3)}{2} \Bigg\{  A(p^2_1) \, X_2 
                       +   B(p^2_1) \, X_3   \Bigg\}  - (p_1\leftrightarrow  p_2)   \ .                                  
\label{EQ:lambda_4} 
\end{eqnarray}
\end{small}
$\!\!\!\!\!$
These expressions comply with the symmetries under interchange of quark momenta listed in (\ref{Eq:VertexDiscretas}).
Moreover, as discussed in \cite{Oliveira:2018fkj} all $\lambda_i$ are regular functions.

The tree level perturbative solution of QCD for the quark-ghost kernel form factors gives $X_0  = 1$ and $X_{1,2,3} = 0$ \cite{Davydychev:2000rt}
and, in this case, the longitudinal quark-gluon vertex is uniquely  determined by the quark wave function $Z(p^2)$ and the running quark mass 
$M(p^2)$.
 On the other hand, the non-perturbative solutions of 
QCD \cite{Aguilar:2014lha,Aguilar:2016lbe,Aguilar:2018epe,Aguilar:2018csq,Oliveira:2018ukh} 
result in $X_{1,2,3}$ that deviate significantly from their tree level values, specially at infrared mass scales. 
These non-per\-tur\-ba\-ti\-ve solutions for $X_0$ return a function that hardly deviates from its tree level perturbative value
\cite{Rojas:2013tza,Aguilar:2014lha,Aguilar:2018csq,Oliveira:2018ukh,Serna:2018dwk} and $X_0 \approx 1$ seems to be a good approximation for this form factor.
We call the reader's attention that so far we only have approximate non-perturbative solutions for QCD.
At a qualitative level, the various approaches seem to produce compatible results. 
See also \cite{Pelaez:2015tba,Binosi:2016wcx,Sultan:2018qpx,Campagnari:2019zso} for other recent studies of  the quark-gluon vertex.

There are now results from  full QCD lattice simulations for $\lambda_1(p^2)$  in the soft-gluon limit \cite{Oliveira:2016muq}.
This information  should be incorporated in the inversion of the gap equation to compute the quark-gluon vertex.
The longitudinal form factors $\lambda_1$ in this limit \cite{Oliveira:2018fkj} is given by
\begin{small}
\begin{equation}
   \lambda_1 (p^2) = \frac{F(0)}{Z(p^2)} \bigg\{ 1 ~ + ~ 2 \, X_1(p^2) \, M(p^2) ~ + ~ 2 \,  p^2 \, X_3 (p^2) \bigg\} \ ,
   \label{EQ:qgv_soft}
\end{equation}
\end{small}
$\!\!\!\!\!$
where $p$ is the quark momentum. This kinematical configuration is interesting also because it gets no contribution 
from the transverse form factors. This can be easily checked using the transverse basis considered in \cite{Kizilersu:1995iz}.

\section{Decomposing the Schwinger-Dyson Equation}

In the current study we use only the longitudinal components of the quark-gluon vertex; see Eq.~(\ref{EQ:Lvertex})
and Sec. \ref{Sec:formfactors} for detailed expressions for all the form factors.
By taking traces of (\ref{eq:SDE}) one accesses either 
its scalar or vector part.
In order to include inputs from lattice simulations, we perform a Wick rotation in the Schwinger-Dyson equations, writing them in the Euclidean 
space (see e.g. \cite{Oliveira:2018ukh} and references therein) as
\begin{small}
\begin{eqnarray}
  & & B(p^2)   =   Z_2\, m^\mathrm{bm}   \\
   & & \qquad +  ~ C_F Z_1 g^2 \, 
         \int \frac{d^4 q}{(2 \, \pi)^4} ~ \frac{ Z(k^2) \, D(q^2) }{  k^2 + \left[ M(k^2) \right]^2}   \Bigg\{    
              \mathcal{A}_s   +  M(k^2)  \, \mathcal{B}_s
  \Bigg\} ,
 \label{Eq:DSEEucScalar}
 \nonumber \\
 & &  p^2 A(p^2)   =   Z_2\, p^2   \\
 &  & \qquad - ~ C_F Z_1 g^2  \int \frac{d^4 q}{(2 \, \pi)^4} ~ \frac{ Z(k^2) \, D(q^2) }{ k^2 + \left[ M(k^2) \right]^2}  
  \Bigg\{      
  \mathcal{A}_v
    +  M(k^2) \, \mathcal{B}_v \, \Bigg\}\nonumber
\label{Eq:DSEEucVector} 
\end{eqnarray}
\end{small}
with 
\begin{small}
\begin{eqnarray}
& & \mathcal{A}_s(p^2,k^2,q^2) =
           2 ~\lambda_3   ~ \left( p^2 - \frac{(pq)^2}{q^2}  \right)  \nonumber\\
& &  \qquad \qquad \qquad +  ~ \lambda_4   ~ \left[   9 \, (pq) -  2 \,  \frac{(pq)^2}{q^2} - 4 \, p^2 - 3 \, q^2   \right] \, , \\
& &\mathcal{B}_s(p^2,k^2,q^2) =   3 ~\lambda_1 
       +  4 ~\lambda_2   ~\left[  \frac{(pq)^2}{q^2} - p^2  \right]  \, , 
\end{eqnarray}
\begin{eqnarray}
&&\mathcal{A}_v(p^2,k^2,q^2) =
    \lambda_1  ~ \left[  p^2 + 2 \,  \frac{(pq)^2}{q^2} - 3 \, (pq)  \right] \nonumber \\
& &  \qquad \qquad \qquad + ~2 \,  \lambda_2  
                       ~ \bigg[ p^2 q^2 + 2 \, \frac{(pq)^3}{q^2} \nonumber \\
                 &  & \qquad \qquad \qquad - ~\left( 2 \, \frac{p^2}{q^2} + 1 \right) (pq)^2  - 2 \, p^2 (pq) + 2 \, p^4 \bigg] , 
\end{eqnarray}
\begin{eqnarray}
&&  \mathcal{B}_v(p^2,k^2,q^2) = 
          2 ~ \lambda_3   \left[  p^2 - \frac{(pq)^2}{q^2}  \right]  \nonumber \\
          & & \qquad \qquad \qquad \qquad + ~ \lambda_4   \left[  4 \, p^2 + 2 \, \frac{(pq)^2}{q^2}  - 3 \, (pq) \right]   , 
\end{eqnarray}
\end{small}

\noindent
where $k = p -q$, $C_F = 4/3$ is the Casimir invariant associated to the SU(3) fundamental representation
and we used $\lambda_i\equiv\lambda_i (p^2 , \, k^2, \, q^2)$ to simplify the notation.

\subsection{Results from 1-loop perturbation theory} 

The  full set of $\lambda_i$'s were computed  in \cite{Bermudez:2017bpx} in perturbation theory to one-loop for arbitrary linear covariant gauges. 
The authors  report full expressions for the various form factors at several kinematic configurations.
For the symmetric limit where $p^2 = k^2 = q^2$, the leading behaviour for the non-vanishing form factors are
\begin{equation}
 \lambda_1 (p^2) ~ \propto ~ 1 \ , \qquad
  \lambda_2 (p^2)\,\, \text{and} \,\,\lambda_3 (p^2)  ~ \propto ~ 1 / p^2 
  \label{EQ:PertScalings}
\end{equation}
up to logarithmic corrections. The  solution of the STI for $\lambda_1$   reads
\begin{eqnarray}
   & & \lambda_1 (p^2) = F(p^2) \Bigg\{ A(p^2) \bigg[ X_0  + \frac{5}{2} \, p^2 \, X_3 \bigg] + 2 \, B(p^2) \, X_1  \Bigg\} \ \nonumber\\
   && \quad \mbox{ and }~ X_2 = X_1 \, .
\end{eqnarray} 
At large quark momentum  $\lambda_1 (p^2)$, $F(p^2)$, $A(p^2)$ and $B(p^2)$  become constants and this expression suggests that 
$X_0 \sim 1$, $X_1 \sim 1$ and $X_3 \sim 1/p^2$. 
These  are minimal scaling laws in the sense that some of the form factors can behave as
larger negative power of $p^2$.
A similar analysis of the symmetric limit for $\lambda_2$ and $\lambda_3$ returns minimal scaling laws compatible with
the above ones. 
The tree level expressions for the quark-ghost kernel form factors are $X_0 = 1$ and $X_{1, 2, 3} = 0$ and these results
should be recovered in the UV limit. This suggests that the minimal scaling law for $X_1$ is not that previously reported but, instead, 
a larger negative  power of $p^2$ as e.g. 
\begin{equation}
X_0 \sim 1 ~ , \qquad X_1 \sim 1/p^2 \qquad\mbox{ and }\qquad  X_3 \sim 1/p^2  \ .
\label{EQ:MaisumaScalingLaw}
\end{equation}
This results are in good agreement with the scaling analysis performed in \cite{Oliveira:2018ukh} and are compatible with the
minimal scaling laws derived directly from one-loop perturbation theory quoted above.

\subsection{The Longitudinal Quark-Gluon Vertex \label{Sec:formfactors}}

In order to compute a solution for vertex from the SDE we assume functional forms for the $X_i$'s. 
As in \cite{Oliveira:2018ukh}  the gluon, the ghost \cite{Duarte:2016iko} and the quark \cite{Oliveira:2018lln}  
propagators are taken from lattice simulations and modelled to reproduce its correct UV perturbative behaviour. Exact
expressions can be found in App. \ref{Sec:propagators}. Moreover, as discussed in \cite{Oliveira:2018ukh}, 
following the analysis of the lattice data for soft-gluon limit  performed in \cite{Oliveira:2018fkj}, an ansatz for the quark-ghost form
factors is
\begin{eqnarray}
X_0 (p^2, k^2, q^2) & \equiv & X_0(q^2) , \label{Eq:AnzX0} \\
X_1 (p^2, k^2, q^2) & = &  X_2 (p^2, k^2, q^2) \equiv D\left( \frac{p^2 + k^2}{2} \right) ~ Y_1(q^2) ,  \label{Eq:AnzX1}  \nonumber \\
\\
X_3 (p^2, k^2, q^2) & \equiv  & D\left( \frac{p^2 + k^2}{2} \right) ~ Y_3(q^2)  \ . \label{Eq:AnzX3} 
\end{eqnarray}
This ansatz solves the SDE with a relative error smaller than 4\% - see \cite{Oliveira:2018ukh} for further details.

The parameterisations (\ref{Eq:AnzX0}) - (\ref{Eq:AnzX3}) have to comply with the exact results derived in \cite{Aguilar:2014lha}
\begin{equation}
   X_0 (p^2, \, p^2, \, 0) = 1 ~ \mbox{ and } ~
   X_1(p^2, \, p^2, \, 0 ) = X_2(p^2, \, p^2, \, 0) \  .
   \label{Eq:Binosi}
\end{equation}
The second condition is already built in the ansatz. Further, 
the ansatz (\ref{Eq:AnzX0}) - (\ref{Eq:AnzX3})
 should  be compatible with the scaling laws Eqs. (\ref{EQ:PertScalings}) and (\ref{EQ:MaisumaScalingLaw}). 
Recall that the gluon propagator $D(q^2)$ scales as $1/q^2$ at large $q^2$ and, therefore, $Y_1 (q^2)$ and $Y_3 (q^2)$ should approach a 
constant or scale, at large $q^2$, with a negative power of $q^2$ for large gluon momentum.

The explicit expressions for the longitudinal quark-gluon vertex form factors are derived
from Eqs.(\ref{EQ:lambda_4}), (\ref{Eq:AnzX0}), (\ref{Eq:AnzX1}) and (\ref{Eq:AnzX3}). 
We can write the longitudinal components of the form factors as
\begin{small}
\begin{eqnarray}\label{FFLONG}
 &&\lambda_1  =   \frac{F(q^2)}{2} \Bigg\{  
  2 \, B(p^2) \, D\left( \frac{p^2 + k^2}{2} \right) ~ Y_1(q^2)   \nonumber \\
              && \quad
              +A(p^2) \left[ X_0(q^2) + \left( p^2 - p \cdot k \right) D\left( \frac{p^2 + k^2}{2} \right) ~ Y_3(q^2) \right]  \Bigg\} \nonumber \\ && \quad + (p \leftrightarrow  k) \ ,
  \nonumber \\
 &&\lambda_2   = 
               \frac{F(q^2)}{2 (k^2 - p^2)} 
               A(p^2)\! \Bigg\{\left( p^2 + p\! \cdot k \right) D\left( \frac{p^2 + k^2}{2} \right) ~ Y_3(q^2)  
               \nonumber \\ 
               && \quad - X_0(q^2) \Bigg\}  + (p\leftrightarrow  k) \ ,
 \nonumber \\
&& \lambda_3   =  
                   \frac{F(q^2)}{p^2 - k^2}    \Bigg\{ 
                    A(p^2) \left[ p^2  + p \cdot k \ \right]  D\left( \frac{p^2 + k^2}{2} \right) ~ Y_1(q^2) \nonumber \\
                   && \quad  +    B(p^2) \, X_0(q^2) 
                   \Bigg\}+ (p\leftrightarrow  k)  \ ,
           \nonumber      \\
&& \lambda_4   =   - \frac{F(q^2)}{2} D\left( \frac{p^2 + k^2}{2} \right) \Bigg\{  A(p^2) ~ Y_1(q^2) +   B(p^2)  ~ Y_3(q^2)   \Bigg\}
 \nonumber \\ && \quad - (p\leftrightarrow  k)   \, ,
\end{eqnarray}
\end{small}

\noindent
where $A(p^2)=1/Z(p^2)$, $B(p^2)=M(p^2)/Z(p^2)$ with $Z(p^2)$ from Eq. (\ref{quark_Z_function}) and $M(p^2)$ given in Eq. (\ref{Eq:RunningMass}). 
The form factors
 $X_0 ( q^2)$, $Y_1(q^2)$ and $Y_3(q^2)$ are given later by Eq.~(\ref{PADE_X3}). 
As written explicitly 
in Eqs.~(\ref{FFLONG}), all the symmetries coming from 
charge conjugation invariance \cite{Davydychev:2000rt}
written in (\ref{Eq:VertexDiscretas}) are satisfied by our longitudinal form factors ansatz. 

 The use of the above ansatz with the gluon propagator as given in Eq. (\ref{Eq:global_gluon_Fit}), of the ghost propagator as in Eq. (\ref{Eq:global_ghost_Fit}) and of
the quark functions given in (\ref{quark_Z_function}) and (\ref{Eq:RunningMass}) together with the longitudinal quark-gluon vertex (see   Sec. ~\ref{Sec:formfactors}) and the exact solution
 (\ref{EQ:lambda_4}) of the Slavnov-Taylor identity (\ref{EQ:STI}) allows for the computation of a solution of the Schwinger-Dyson Equations, whose
 results will be discussed in the following sections.

\section{Incorporating the Soft-Gluon Limit}

To take into account the full QCD simulations results for the soft-gluon limit of  $\lambda_1$ reported in
\cite{Oliveira:2016muq} we insert (\ref{Eq:AnzX0}) - (\ref{Eq:AnzX3}) into (\ref{EQ:qgv_soft}) obtaining, after rotation to the Euclidean,
\begin{small}
\begin{equation}
   \lambda_1 (p^2) = \frac{F(0)}{Z(p^2)} \bigg\{ 1 ~ + ~ 2 \, M(p^2) \, D(p^2)  \, Y_1(0) ~ - ~ 2 \,  p^2 \,  D(p^2) \, Y_3 (0) \bigg\} \ .
   \label{EQ:qgv_soft_Anz}
\end{equation}
\end{small}

\begin{figure}[t]
   \centering
   \includegraphics[scale=0.35]{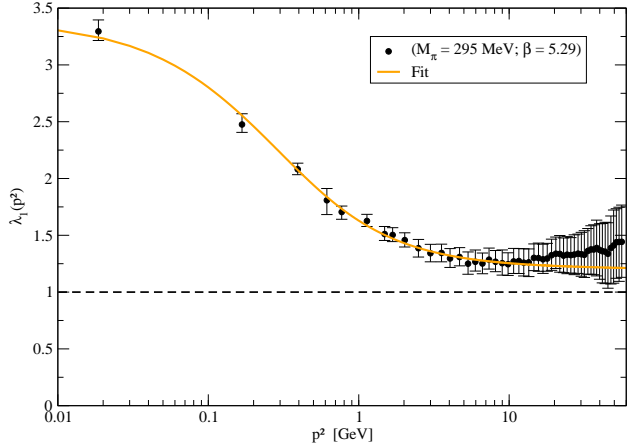}
   \caption{Lattice data for $\lambda_1$ from full QCD simulations \cite{Oliveira:2016muq}, in lattice units, and the fit to (\ref{EQ:qgv_soft_Anz}). }
   \label{fig:fit}
\end{figure}

\noindent
A correlated fit of (\ref{EQ:qgv_soft_Anz}) to the lattice data using the ensemble with a pion mass of  295 MeV and a $\beta = 5.29$
\cite{Oliveira:2016muq},
relying on the expressions for $D(p^2)$ and $M(p^2)$ given in \cite{Oliveira:2018ukh}, see also App. \ref{Sec:propagators}, returns
$Y_1(0) = 0.1726 \pm 0.0074$ GeV,  $Y_3 (0)  = -0.0806 \pm 0.008944$ and $F(0) / Z(0)  = 1.211 \pm 0.029$ for
a  $\chi^2/d.o.f. = 0.15$. 
In Fig.  \ref{fig:fit} we show both the lattice data and the fitted function (\ref{EQ:qgv_soft_Anz}).
The above numbers for $Y_1(0)$ and $Y_3(0)$ provide a normalisation at zero momentum, that depends on the renormalisation scale $\mu$, 
for  $Y_1(q^2)$  and $Y_3(q^2)$. 

For the comparison of the fitted value for the ratio $F(0) / Z(0)$ with the expression given in  \cite{Oliveira:2018ukh} and reproduced in App. \ref{Sec:propagators}
one has to consider the normalisation factor that comes from the renormalisation of $\lambda_1$. 
The fit was performed using bare lattice data and a gluon and ghost
propagators renormalised in MOM-scheme at $\mu = 3$ GeV. This should be corrected before  doing any comparison.
One can estimate this global normalisation factor for $\lambda_1$ from the plot by demanding that $\lambda_1 (\mu^2) = 1$.
The quark renormalisation constant can be computed after setting $Z(\mu^2) = 1$. These renormalisation constants allows 
the rescaling of the fitted ratio $F(0) / Z(0)$ and only then estimate the value of $F(0)$ that corresponds to a 
renormalised ghost propagator $F( \mu^2 ) = 1$.  It is this value, for $\mu = 3$ GeV, that should be compared with the ghost fitting 
function at zero momentum considered in \cite{Oliveira:2018ukh}. 
It turns out that the fitted ghost propagator function used in \cite{Oliveira:2018ukh} and that computed from the fit to the soft-gluon limit
 are compatible within one standard deviation. This gives us further confidence that the ansatz 
(\ref{Eq:AnzX0}) - (\ref{Eq:AnzX3}) for quark-ghost kernel form factors is able to capture the essential of the QCD dynamics.

\section{Inverting the Schwinger-Dyson Equations}

\begin{figure}[t] 
   \centering
   \includegraphics[width=3.5in]{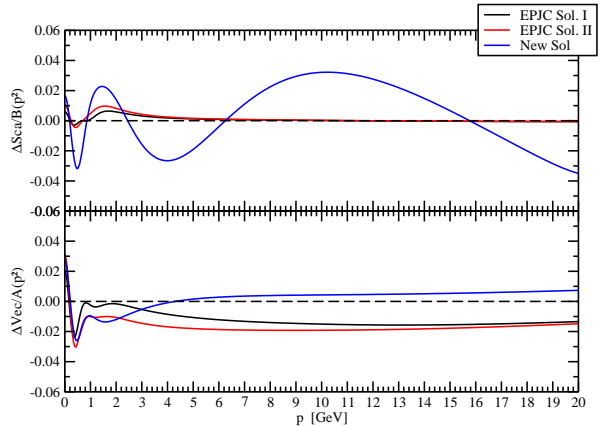}  
   \caption{Relative error for the Schwinger-Dyson equations for the solutions I and II reported in \cite{Oliveira:2018ukh} for $\alpha_s = 0.22$
                and the new solution considered here, computed using Pad\'e approximants and taking into account the soft-gluon limit. In both solutions
                the propagators were renormalised at $\mu = 4.3$ GeV using the MOM-scheme.}
   \label{fig:erro}
\end{figure}

\begin{figure}[h] 
   \centering
   \includegraphics[width=2.2in]{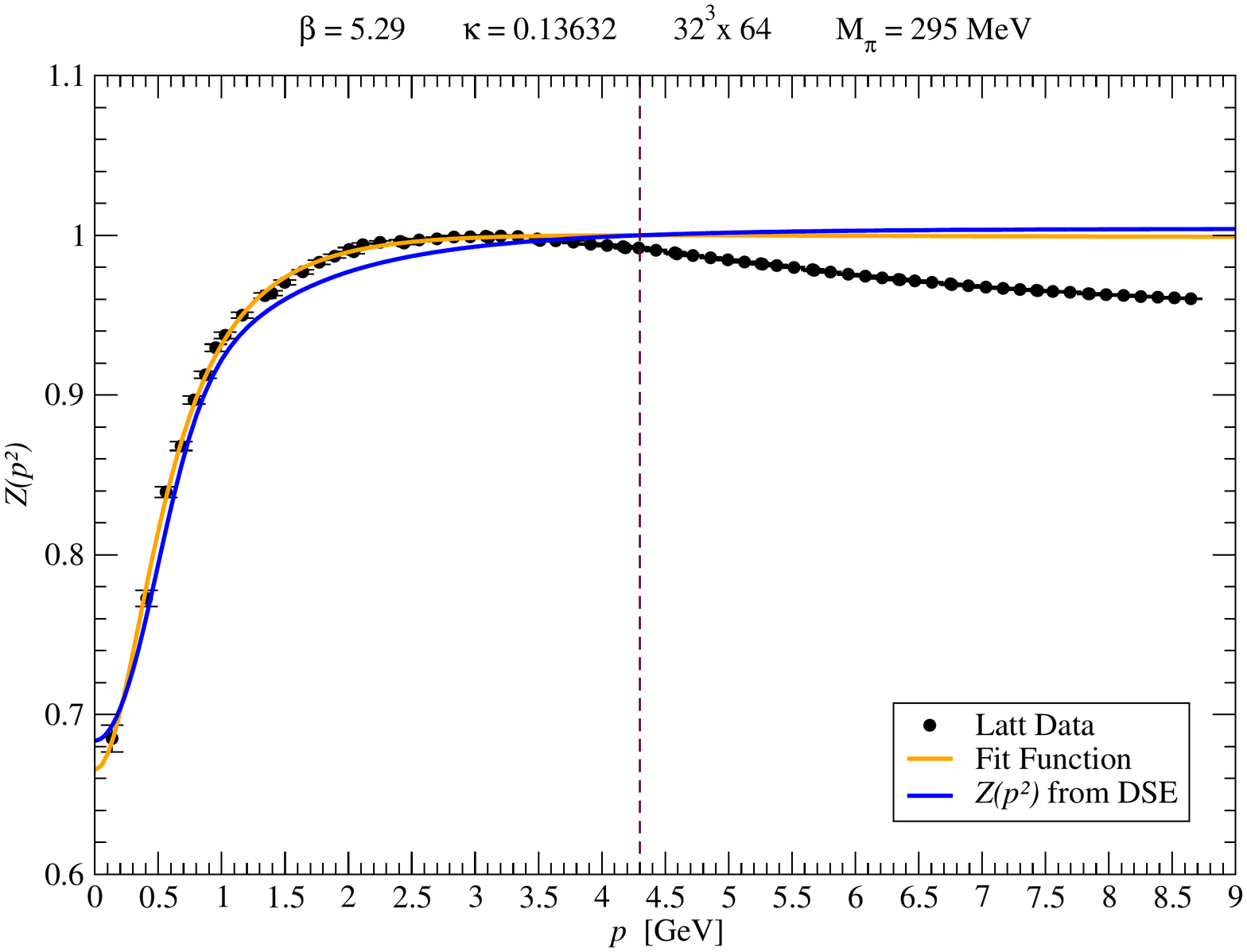}  
   \includegraphics[width=2.2in]{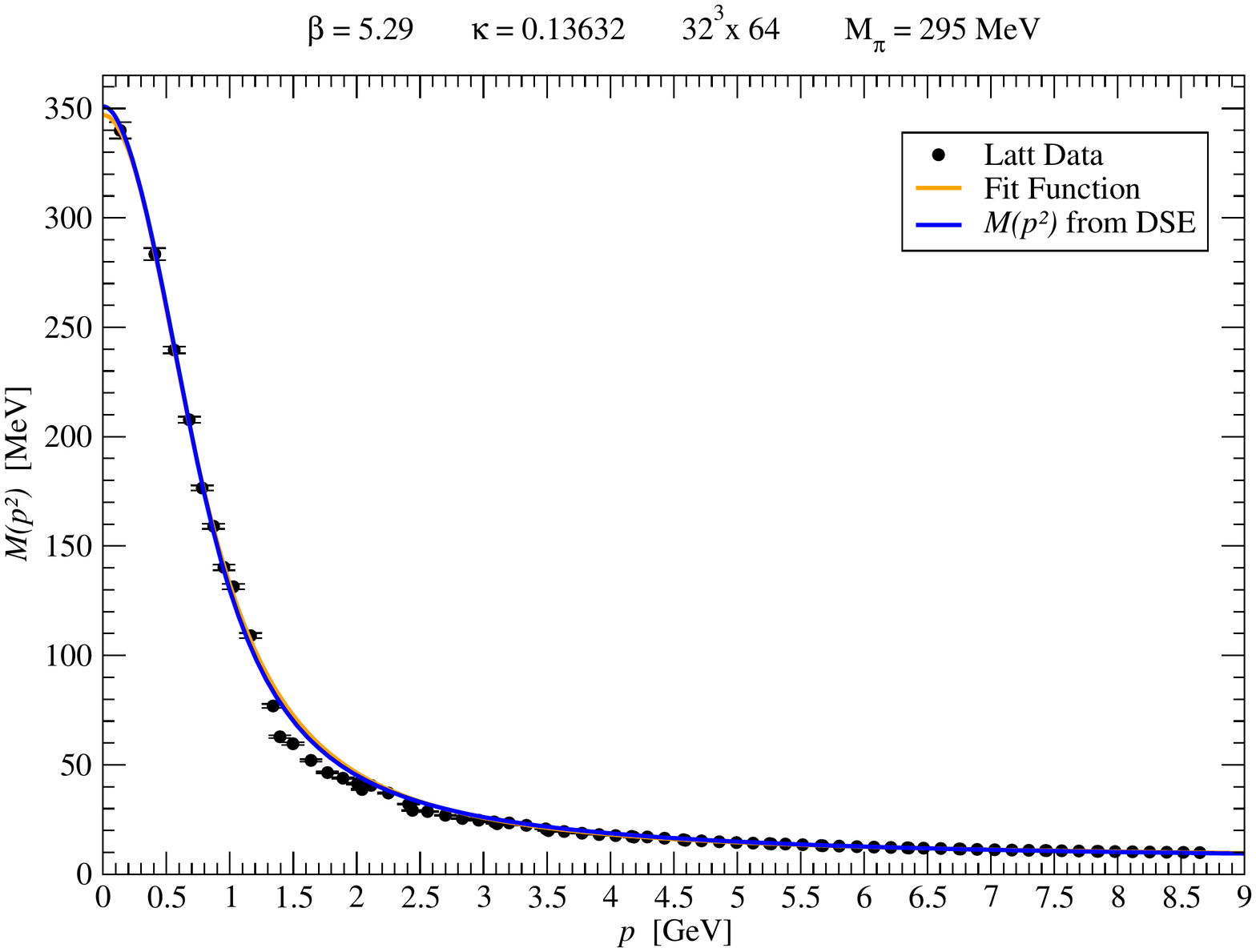}   \\
   \caption{The quark wave function and the running quark mass as given by the r.h.s. of the Schwinger-Dyson equations for our vertex solution,
                 compared with the fit functions that parametrise $Z(p^2)$ and $M(p^2)$ and the lattice data for the quark propagator. For comparison the
                 quark wave function was renormalised in the MOM-scheme at $\mu = 4.3$ GeV. The lattice data 
                  from $N_f = 2$ full QCD simulation in the Landau gauge~\cite{Oliveira:2016muq,Oliveira:2018lln} for
$\beta = 5.29$, $\kappa = 0.13632$ and for a $32^3 \times 64$ lattice 
                  was rescaled using the parameterisation reported
                 in App. \ref{Sec:propagators}. See \cite{Oliveira:2018ukh} for the discussion on the parameterisation of the lattice quark propagator.}
   \label{fig:DSE_rhs}
\end{figure}

The SDE are a set of two coupled equations that depend, for our ansatz, on three form factors.
Furthermore, we would like to take into account the normalisation for $X_0$ given in Eq. (\ref{Eq:Binosi}),
the corresponding UV limit and the normalisations for $Y_1(0)$ and $Y_3(0)$  from the fit to the soft-gluon limit of the lattice data, 
after appropriate rescaling to comply with the renormalisation scale. We recall the reader that in \cite{Oliveira:2018ukh} we used
$\mu = 4.3$ GeV as renormalisation scale and it will also be used here to solve the SDE.
All these constraints can be taken in the calculation if all the functions are parametrised by Pad\'e approximants
\begin{eqnarray}
 X_0 ( q^2)  & =  & \frac{ 1 + a_{02} \, q^2 + a_{04} \, q^4}{1 + b_{02} \, q^2 + a_{04} q^4} \ ,  
 \nonumber \\
    Y_1(q^2) & = & \frac{Y_1(0) + a_{12} \, q^2 + a_{14} \, q^4  + a_{16} \, q^6 + a_{18} \, q^8}
                                    { 1 + b_{12} \, q^2 + b_{14} \, q^4 + b_{16} \, q^6 + b_{18} \, q^8} \  , 
                                    \nonumber\\
    Y_3(q^2) & = & \frac{Y_3(0) + a_{32} \, q^2 + a_{34} \, q^4  + a_{36} \, q^6 + a_{38} \, q^8}
                                    { 1 + b_{32} \, q^2 + b_{34} \, q^4 + b_{36} \, q^6 + b_{38} \, q^8} \  . \nonumber\\ \label{PADE_X3}
\end{eqnarray}
The coefficients in  (\ref{PADE_X3}) were computed relying on simulating annealing to minimize the sum of the 
relative error of the scalar and vector equations. The numerical experiments show that it is relative easy to produce ``solutions'' whose maximum
relative error for the SDE is of the order of 15\%. However, for errors below the 10\% value we found a single solution. 
As seen in Fig. \ref{fig:erro} we found a solution that solves the SDE equations with a relative error, on each equation, below the 4\% level.
In the minimisation and to avoid poles on the Euclidean momenta real axis it was assumed that all the coefficients in the denominator are 
positive real numbers. We also report in Fig.~\ref{fig:DSE_rhs} the quark wave function and the running quark mass, computed using
our vertex solution, for the r.h.s. of the Schwinger-Dyson equations.

Our parametrisation for $X_0(q^2)$ is the simplest Pad\'e approximant that is compatible with the normalisation conditions $X_0(0) = 
X_0( + \infty ) = 1$ and allows for small deviations from unity as found in previous 
investigations \cite{Oliveira:2018ukh,Aguilar:2016lbe,Aguilar:2018epe}. Furthermore, taking as guide these previous calculations we expected a 
maximum of $X_0(q^2)$ below 1 GeV. Given that for small $q^2$, the function $X_0(q^2)$ is expected
 to grow, then $b_{02} < a_{02}$. If $X_0(q^2)$ has a maximum above 1 for $q < 1$ GeV, this demands
 $a_{02} < 1$ GeV$^{-2}$. All these constraints for $X_0$ were taking into account in the minimisation process.

In the minimisation of the error we also changed the powers of the numerator and denominator in the Pad\' e approximants for $Y_1(q^2)$ and
$Y_3(q^2)$ but only with those reported above we were able to find a solution of the SDE with a relative error below 4\%. 
During the minimization process we observed that the first function to stabilize was $Y_1(q^2)$, followed by $Y_0(q^2)$ and then by $Y_3(q^2)$. 

In Fig. \ref{fig:erro} we show the relative error for the solution of the
Schwinger-Dyson equations based on Pad\'e approximants and the solutions reported in \cite{Oliveira:2018ukh} computed
with $\alpha_s = 0.22$.
In all cases the relative error is below 4\%.

Fig. \ref{fig:DSE_rhs} shows $Z(p^2)$ and $M(p^2)$ computed from the r.h.s of the SDE when our vertex solution is used. The quark wave
function follows very closely the original parameterisation used as input for the calculation, see App. \ref{Sec:propagators}, with small deviations for
momenta in the range $1 - 3$ GeV but reproducing the lattice data in the infrared region and the correct lattice data for larger momenta - see the discussion
on the usage of the lattice quark propagator data in \cite{Oliveira:2018ukh}. On the other hand, the running quark mass computed from the r.h.s. of the SDE is on top of
the parameterisation for the same function used as input in the calculation.

The results reported in Fig.~\ref{fig:DSE_rhs} help us to understand how the relative errors shown in Fig. \ref{fig:erro} translate into the functions
that describe the quark propagator. We recall that in  Fig. \ref{fig:DSE_rhs} we are not solving the SDE for $Z(p^2)$ and $M(p^2)$ but are performing a
consistency check of our approach and also on the ansatz (\ref{Eq:AnzX0}) - (\ref{Eq:AnzX3}) that we plug in into the solution (\ref{EQ:lambda_4}) of the
Slavnov-Taylor identity. Full expression for the quark-gluon form factors ansatz are given in (\ref{FFLONG}).

\section{Results} 

\begin{figure}[t] 
   \centering
   \includegraphics[width=2.0in]{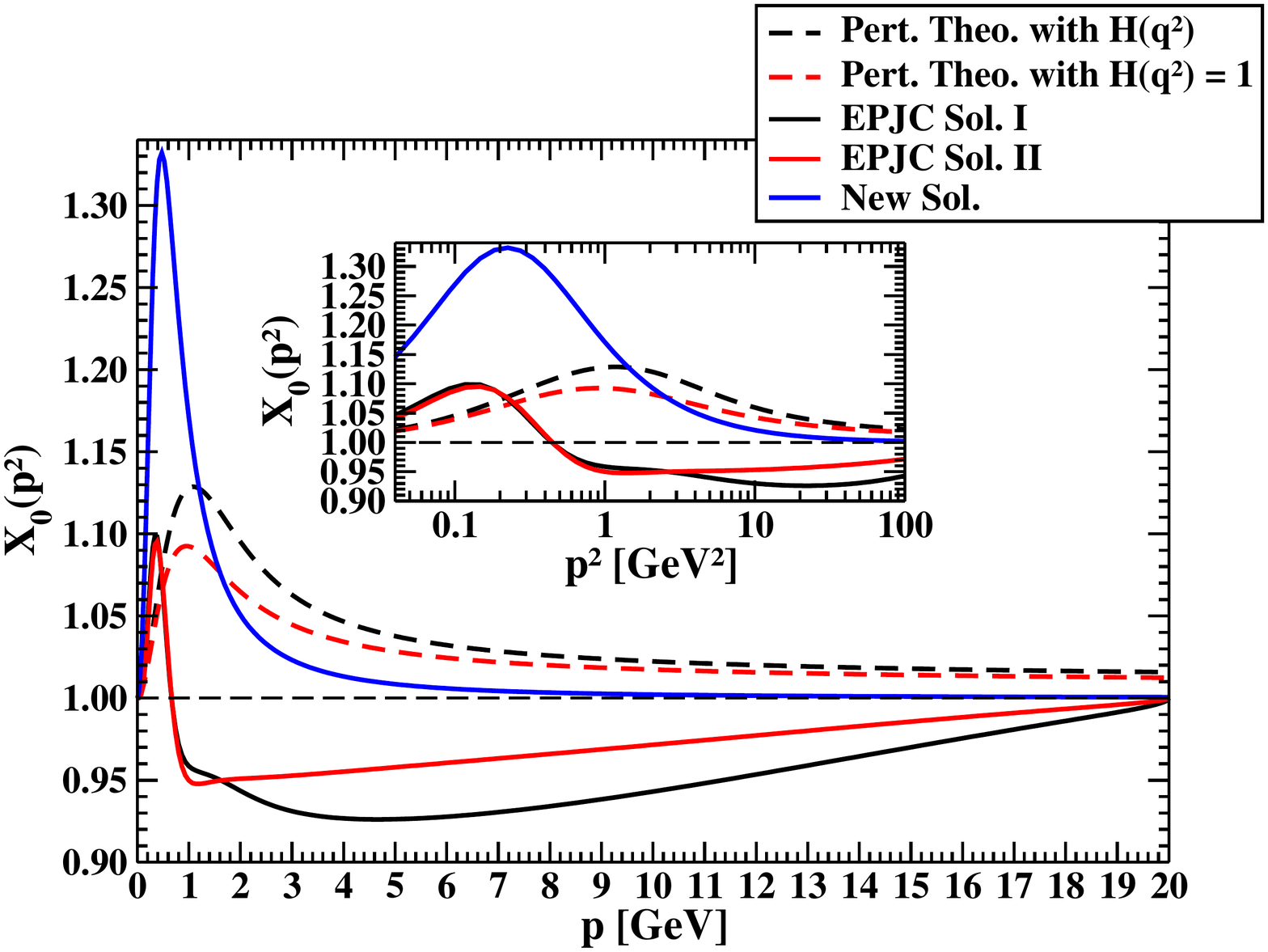}  
   \includegraphics[width=2.0in]{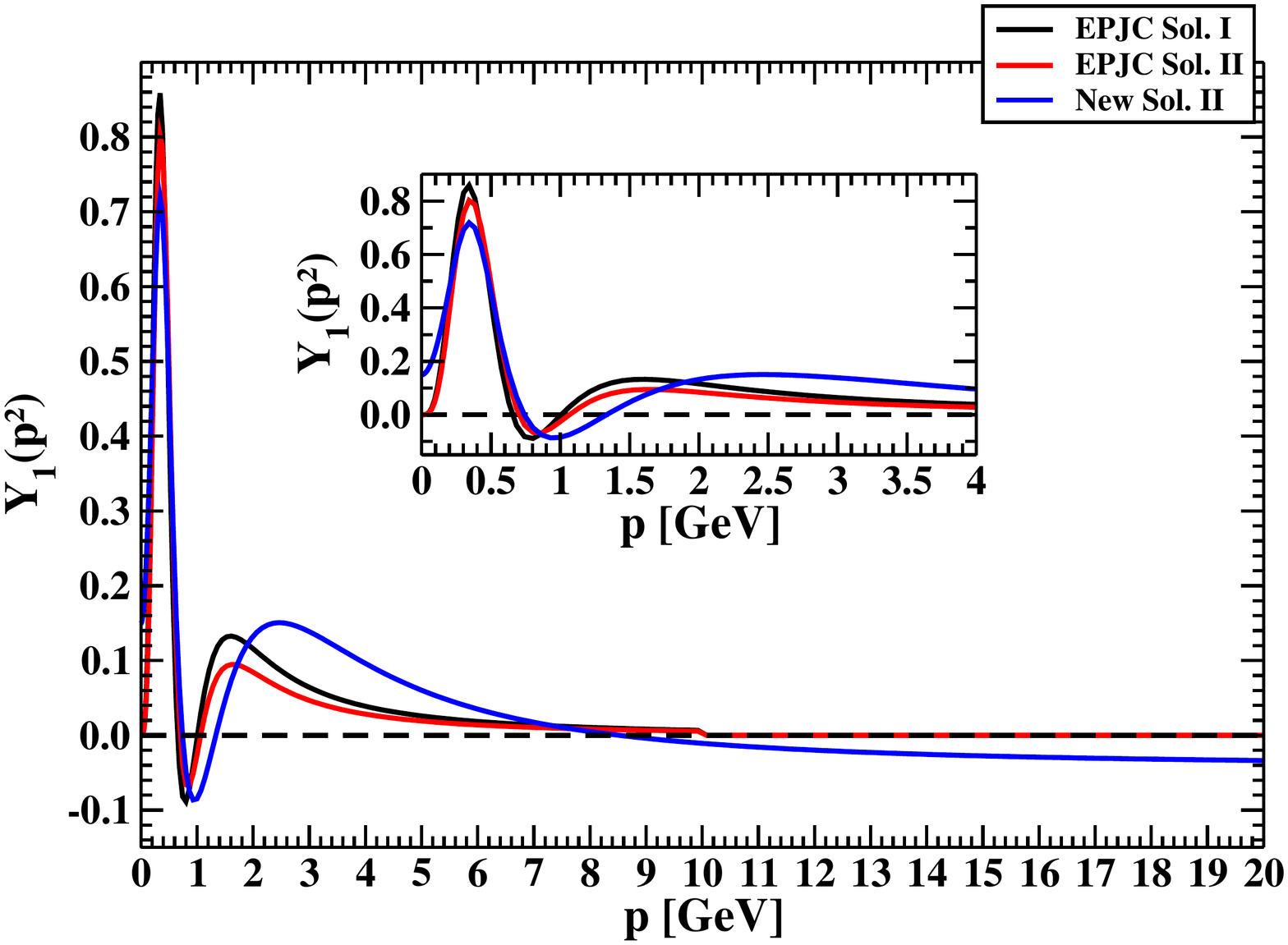}  
   \includegraphics[width=2.0in]{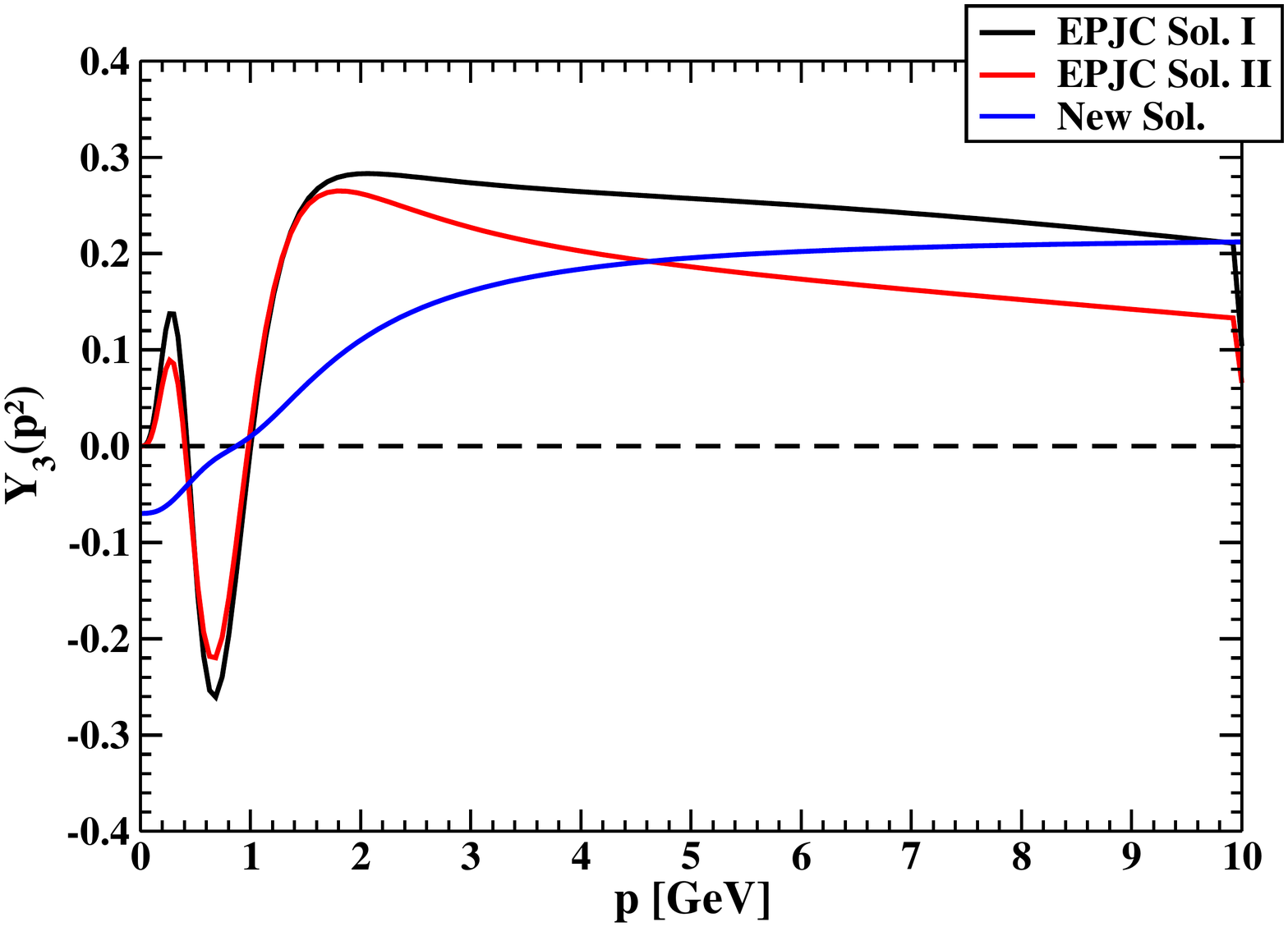}  
   \caption{The solutions I and II found for $X_0(q^2)$, $Y_1(q^2)$ and $Y_3(q^2)$ in \cite{Oliveira:2018ukh} 
                   and the new solution found with the Pad\'e parameterisation. See text for details.
                   Also shown are the 1-loop dressed perturbation theory for $X_0(q^2)$ (dashed lines) using both the tree level gluon-ghost 
                   vertex ($H(q^2)=1$) and an improved vertex ghost-gluon vertex \cite{Dudal:2012zx}.}
   \label{fig:solucoes}
\end{figure}

In Tab. \ref{tab:padefit} the coefficients for the solution that minimise the relative error of the SDE are reported. The corresponding
form factors $X_0(q^2)$, $Y_1(q^2)$ and $Y_3(q^2)$ are shown in Fig. \ref{fig:solucoes} and compared to the solutions computed in
\cite{Oliveira:2018ukh} with a completely different method, where the original SDE were replaced by Tikhonov regularised equations. 
All the represented solutions were computed using the same set of parameters, namely an UV hard cutoff of $\Lambda = 20$ GeV, 
$\alpha_s = 0.22$ and all propagators renormalised at $\mu = 4.3$ GeV. 
Moreover, for the various integrations, angular and momentum, we used exactly the same number of Gauss-Legendre points as in
\cite{Oliveira:2018ukh}.
\begin{table}[t]
   \centering
   \begin{tabular}{l@{\hspace{0.1cm}}|rrrrr} 
      \hline\hline
       $X_0(q^2)$ & 1.00000  & 8.3505  & 20.2310 & & \\
                          & 1.00000  & 3.8906  & 20.2310  & &\\
      \hline\hline
      $Y_1(q^2)$ & 0.14961 & 9.4208 &  -23.2416   & 10.3067 & -0.1379 \\
                          & 1.00000 & 0.005153 & 21.7996 & 15.7754 & 3.2950   \\
      \hline\hline
      $Y_3(q^2)$ & -0.06986 & -1.0907  & 3.8888  & -5.7121 & 3.6823 \\
                          &  1.00000 & 17.4524 & 6.6828 & 19.7787 & 17.7508 \\
      \hline\hline                             
   \end{tabular}
   \caption{Coefficients of the Pad\'e approximant in (\ref{PADE_X3}) in powers of GeV. For each function, the upper line refers to the
   numerator coefficients in increasing power of $q^2$, while in the lower line are the coefficients for the denominator polynomial in
   increasing powers of $q^2$.}
   \label{tab:padefit}
\end{table}

For $X_0(q^2)$ the new solution is enhanced compared to those computed in \cite{Oliveira:2018ukh}, it has a maximum  of $\sim$ 1.35 to be
compared with $\sim$ 1.10 for the old solutions. The maximum of the new solution occurs at slightly larger $q \sim$ 450 MeV for the
Pad\'e based solution and $\sim$ 350 MeV for Tikhonov regularised solutions. The outcome of the one-loop dressed perturbation theory
reported also in Fig. \ref{fig:solucoes} have maxima that are similar to those of the Tikhonov regularised solution but occurring at a much larger scale, 
i.e. for $q \sim$ 1 GeV.
The Pad\'e based solution does not show any minima with $X_0(q^2) < 1$, as seen on the Tikhonov solutions,
and approaches the UV normalisation condition $X_0(+ \infty) = 1$ in a smoother way than the Tikhonov ones. 
In this respect the new solution follows closer the behaviour observed for the predictions of one-loop dressed perturbation theory.

The $Y_1(q^2)$ seen in Fig. \ref{fig:solucoes} are quite similar up to $\sim 1$ GeV. The maximum of the Pad\'e solution being slightly smaller
than those of \cite{Oliveira:2018ukh} and its deep infrared values, i.e. for $q \lesssim 200$ MeV, being  larger to accommodate the lattice 
soft-gluon limit. For $q \gtrsim 1$ GeV, the various curves have similar structures, i.e. the same number of maxima and minima, 
but differ in UV. Here the Pad\'e based solution approaches a negative constant value, while the Tikhonov solutions approach
a positive constant value.

The Pad\'e based solution for $Y_3(q^2)$ is different from those computed in \cite{Oliveira:2018ukh}. It has a simplified
structure that interpolates between its zero momentum value dictated by the lattice soft-gluon limit and a UV constant value that is about the
same found for Sol. I in \cite{Oliveira:2018ukh}.

In \cite{Aguilar:2018epe} the authors solved simultaneously the SDE for the quark propagator together with the quark-ghost kernel, in its
one-loop dressed perturbation theory formulation, to compute the various form factors $X_i$. At the qualitative level, but not quantitatively, 
our results point in the same direction. See \cite{Oliveira:2018ukh} also for notation issues.

\begin{figure*}[h] 
\centering
   \includegraphics[width=1.5in]{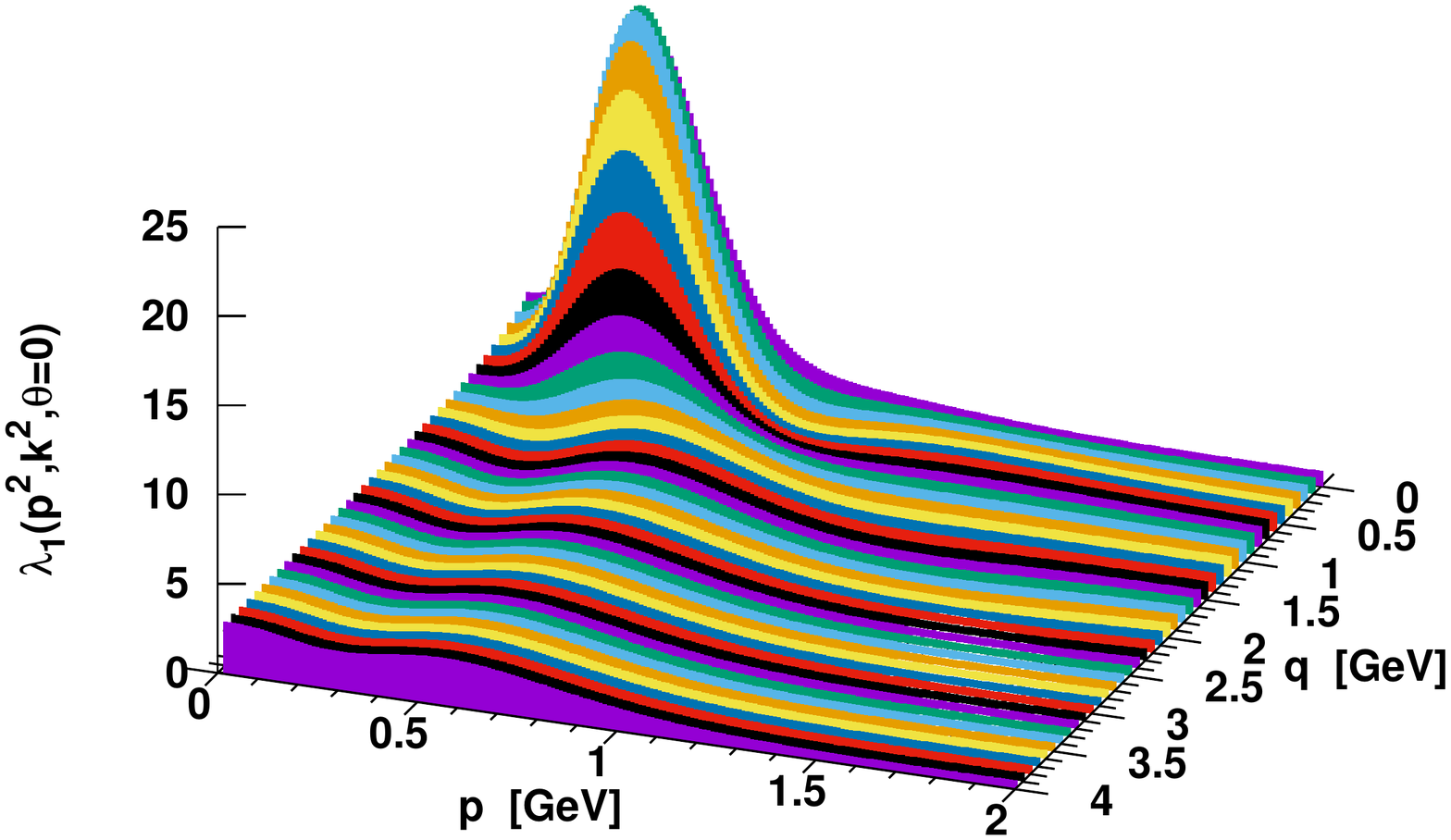} ~
   \includegraphics[width=1.5in]{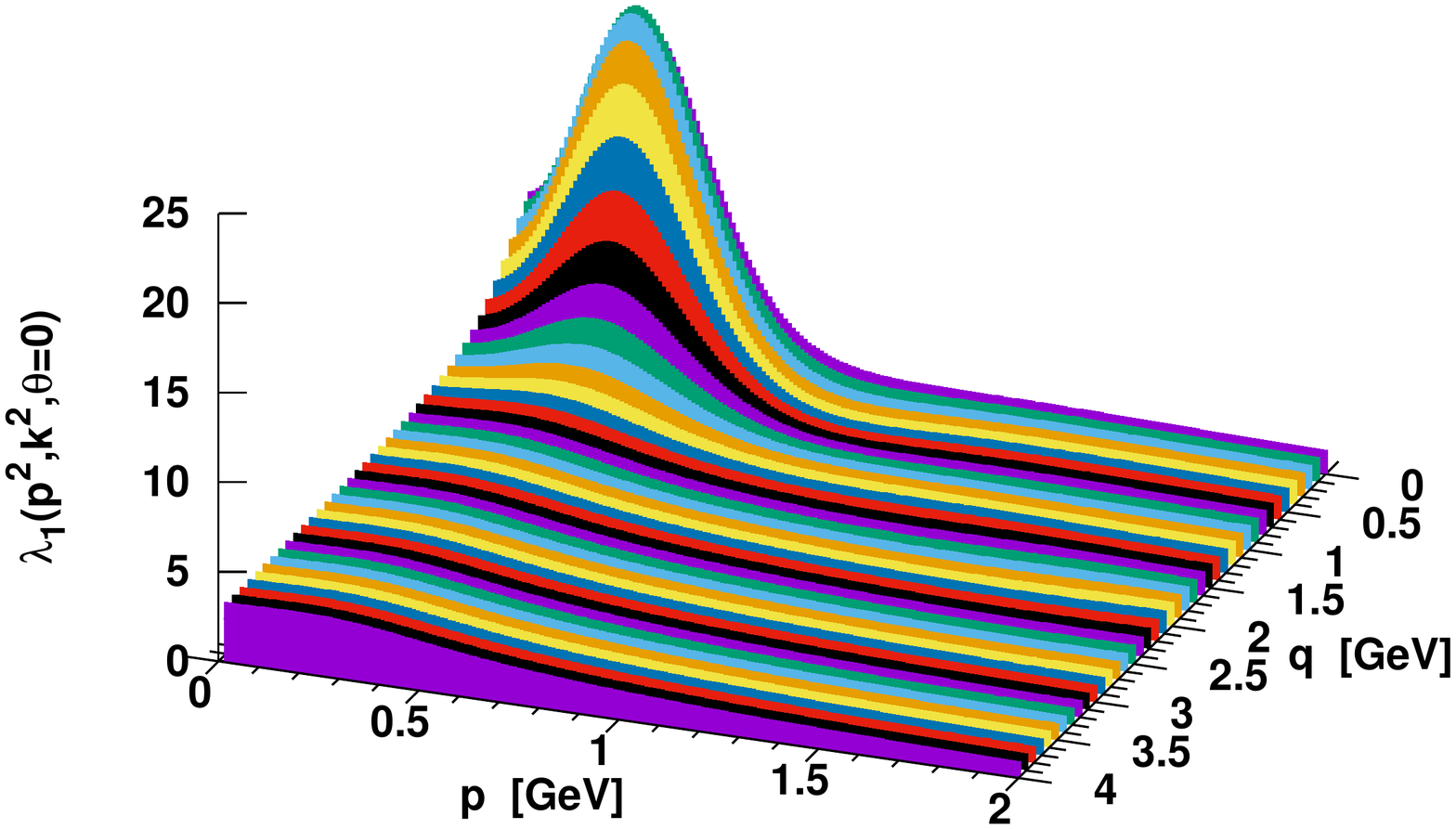}  \\
   \includegraphics[width=1.5in]{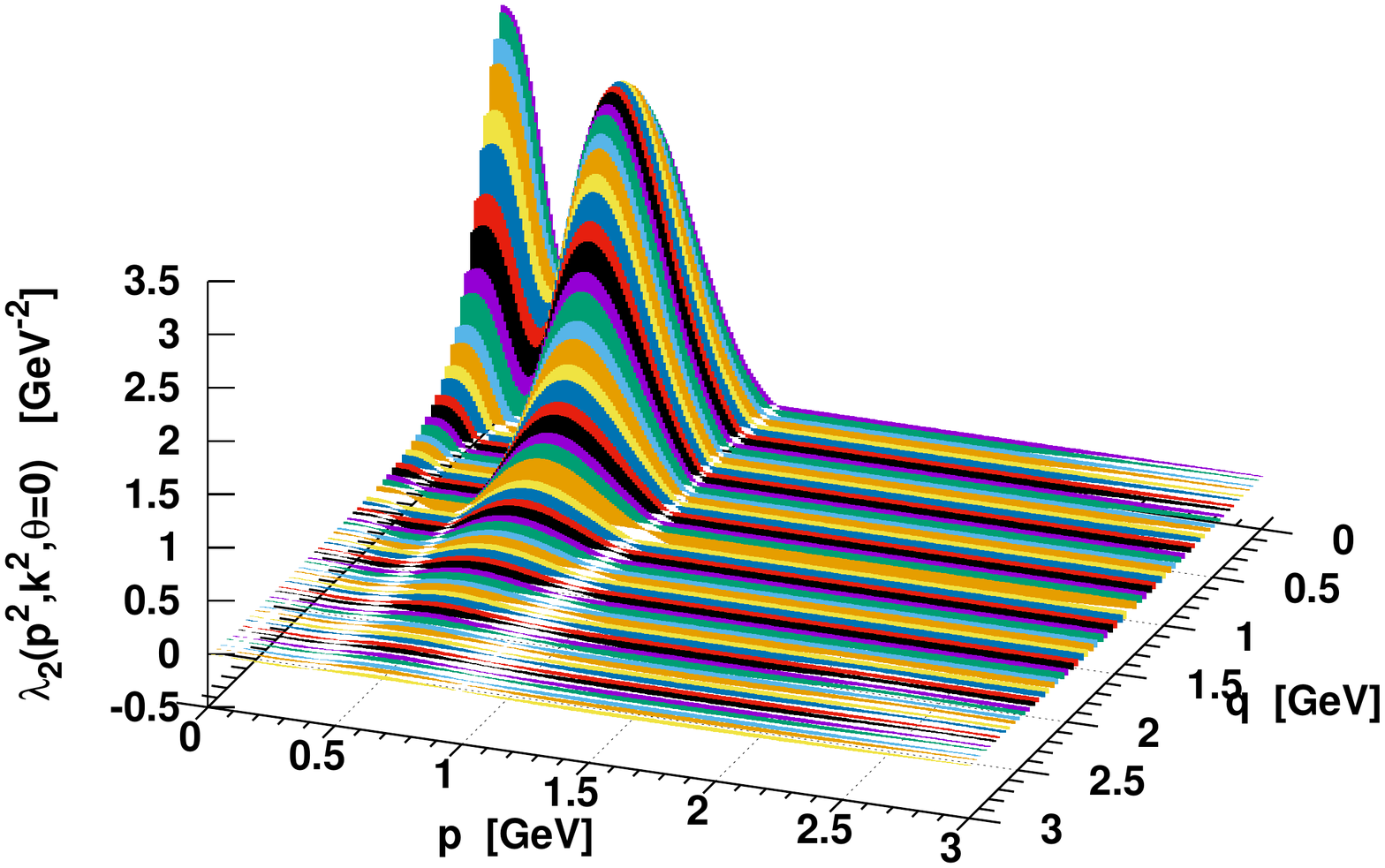} ~
    \includegraphics[width=1.5in]{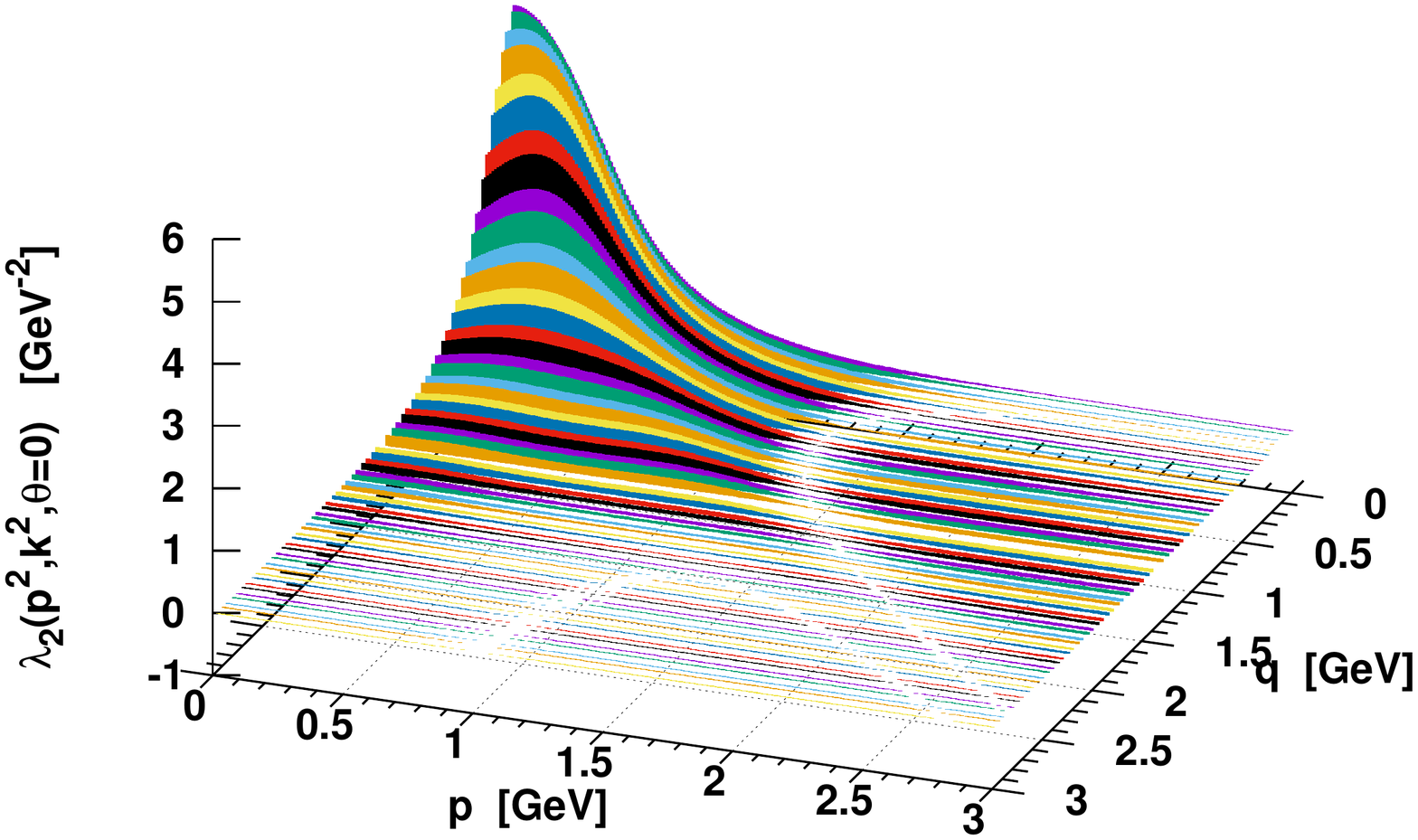}  \\
       \includegraphics[width=1.5in]{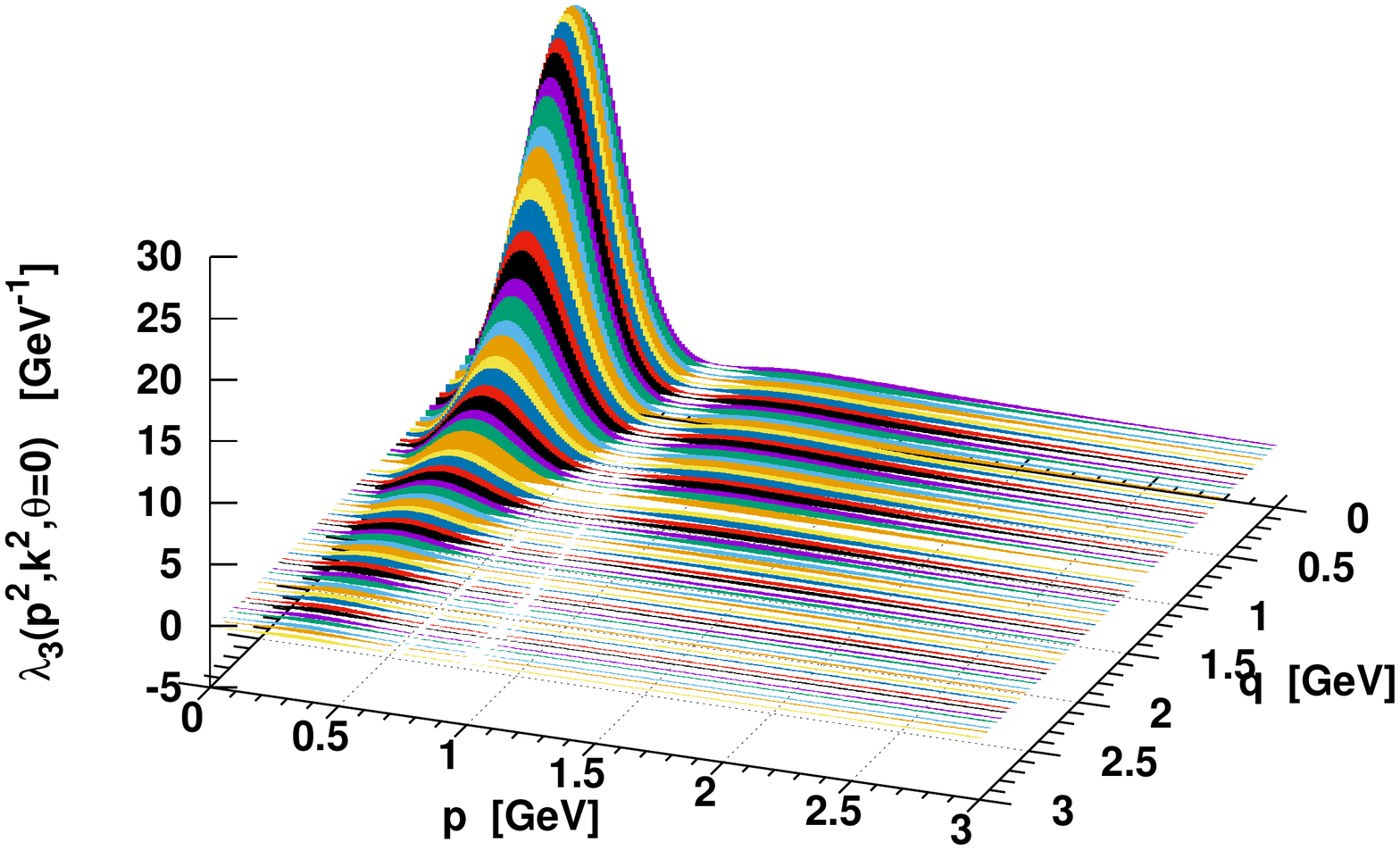} ~
   \includegraphics[width=1.5in]{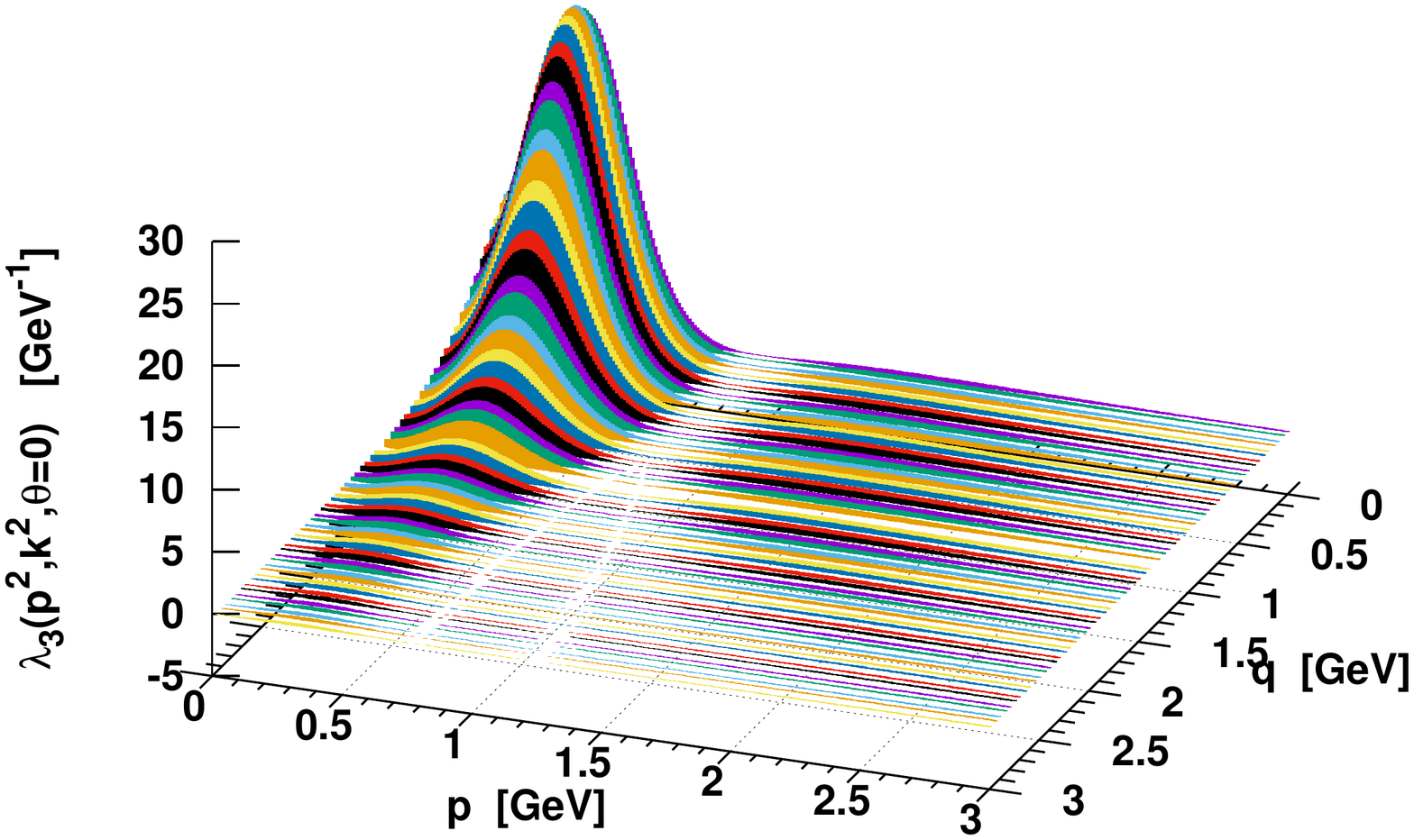}  \\
      \includegraphics[width=1.5in]{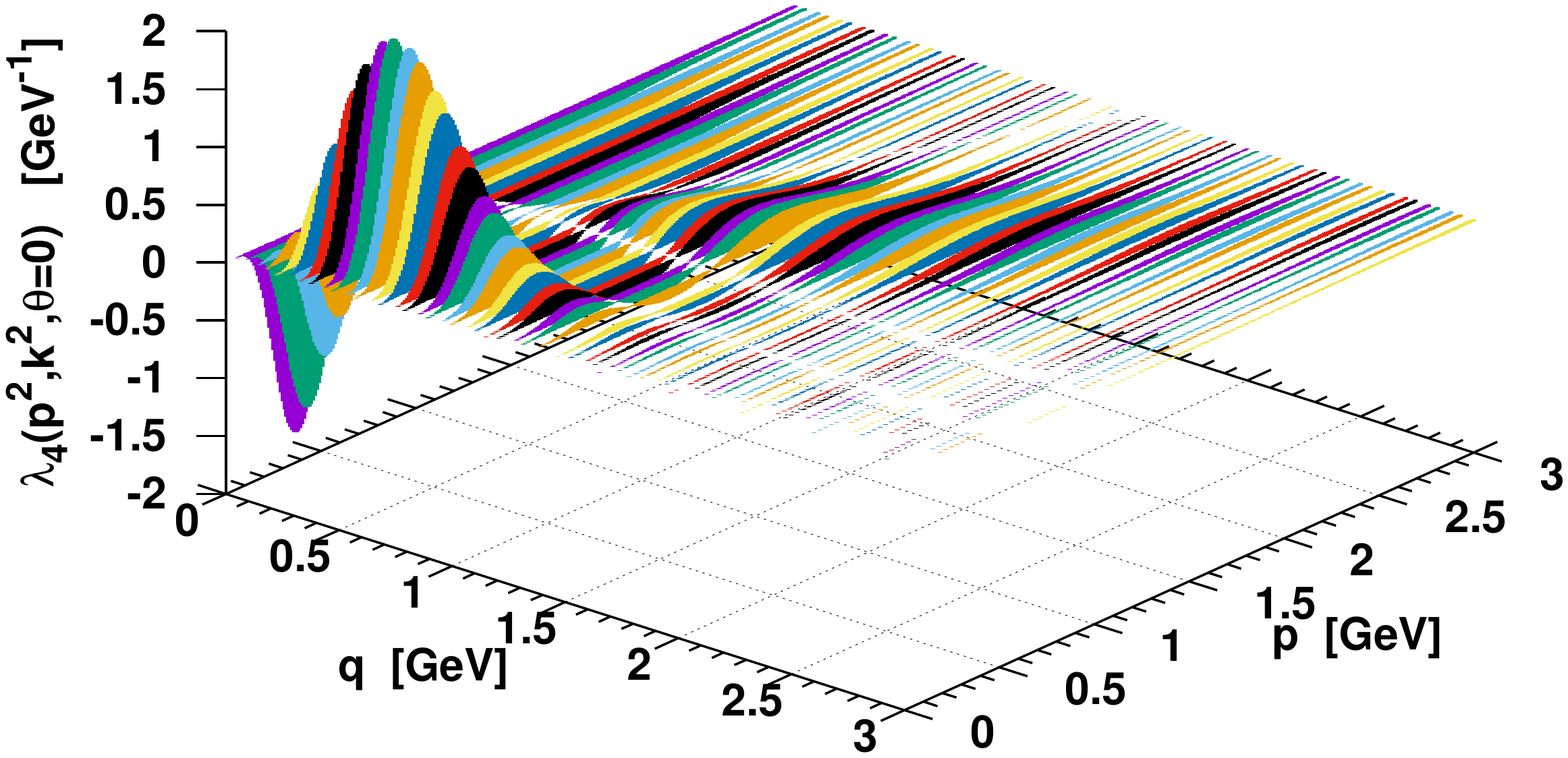} ~
    \includegraphics[width=1.5in]{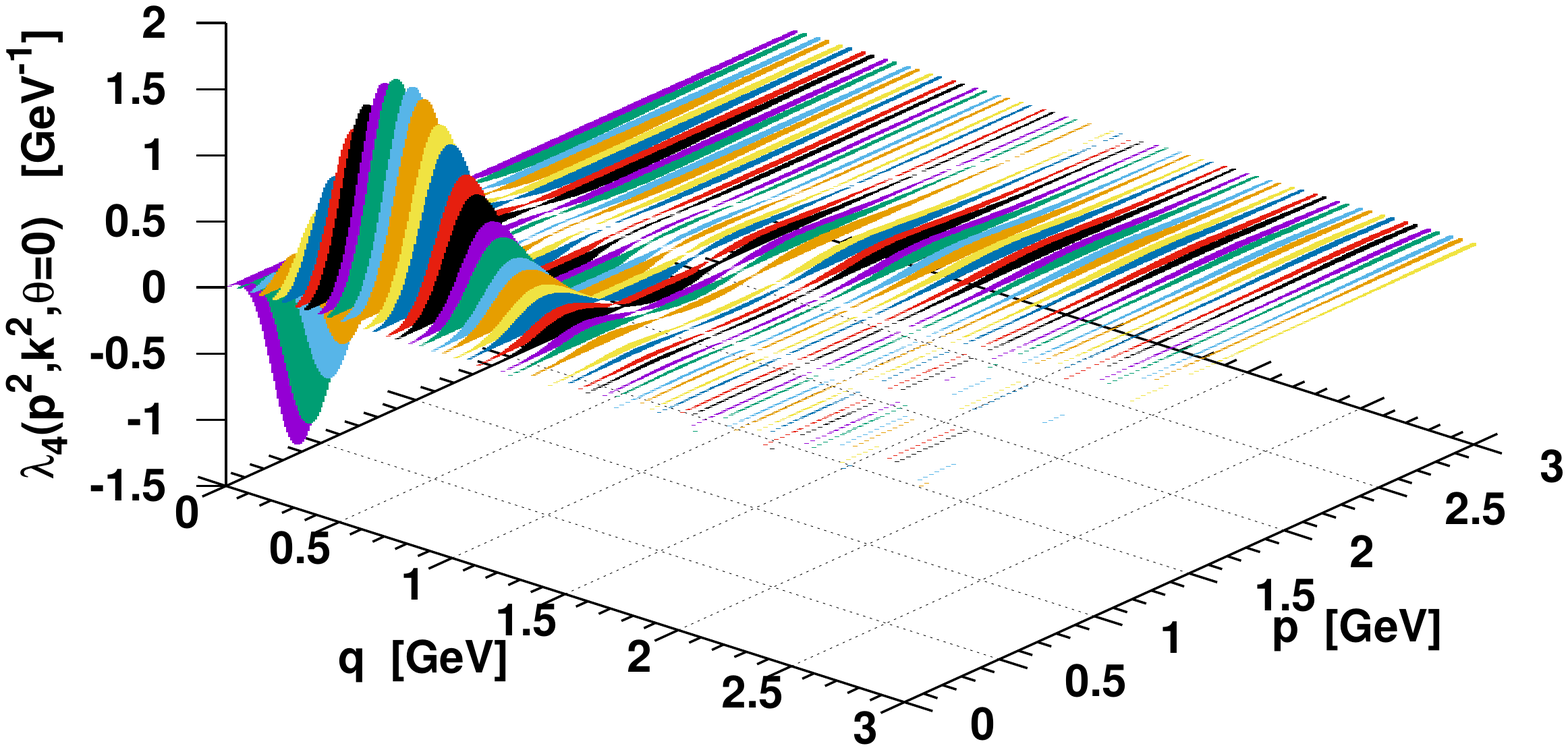} 
   \caption{Form factors for Sol. I reported in  \cite{Oliveira:2018ukh} (left) together with those associated 
                 with the new solution based on Pad\'e approximantions (right).}
   \label{fig:example0}
\end{figure*}

In Fig. \ref{fig:example0} we report the various $\lambda_i$'s, the relevant quark-gluon vertex form factors,
for the new Pad\'e solution and compare them to the corresponding ones for Sol. I computed in \cite{Oliveira:2018ukh}. 
Somehow surprisingly the differences between the two sets of form factors are minimal, with the exception of
$\lambda_2$. These situations  also occurs for other values of $\theta \ne 0$ (not represented here), the angle between the quark and the gluon momentum. 
This  is a welcome feature, as the two solutions were computed in completely different  and independent
ways, giving confidence in our findings. 
In general the $\lambda_i$ based on the Pad\'e solution for $X_0$ and $Y_i$ have slightly less structure as the oscillations observed
in the Tikhonov regularised calculation are not present. 
$\lambda_1$ for the two solutions is very similar with the Pad\'e based calculation showing a clear enhancement in the infrared 
region. 
The scale and the region of the absolute maximum of this form factor is the same for the two solutions. 
A similar conclusion holds also for $\lambda_3$. 
As Fig. \ref{fig:example0} shows (see also the discussion in
\cite{Oliveira:2018ukh}) the main non-perturbative tensor structures of the quark-gluon vertex are associated with $\lambda_1$ and $\lambda_3$ and,
in both cases, their maxima is at $p \sim \Lambda_{QCD}$. 
In what concerns the remaining form factors, the figure also shows that
$\lambda_4$ of the Pad\'e approximation is similar to that published for Sol. I in \cite{Oliveira:2018ukh}. 
A large difference occurs for $\lambda_2$ that is infrared enhanced by about a factor of 2 and shows a single maximum
instead of two for the \cite{Oliveira:2018ukh} solution. For zero momentum, the value of  $\lambda_1 (0)$ in the present calculation is about $2.1$ times the one obtained for 
Sol. I of \cite{Oliveira:2018ukh}. 

Our conclusion being that the inclusion of the information from the lattice soft-gluon limit in the calculation leads to an
infrared enhancement of the quark-gluon vertex. The relative importante of the form factors found in \cite{Oliveira:2018ukh} 
is also observed within the new solution.
Our new calculation supports that the quark-gluon vertex is dominated by the form factors associated to the tree level vertex $\gamma_\mu$
and to $2 p_\mu + q_\mu$, with smaller contributions from the higher order tensor structures $(2 \sla{p} + \sla{q}) (2 p + q)_\mu$ and
$\sigma_{\mu\alpha} (2 p + q)_\alpha$.

\begin{figure}[h]
\centering 
\subfigure{\includegraphics[width=2.0in]{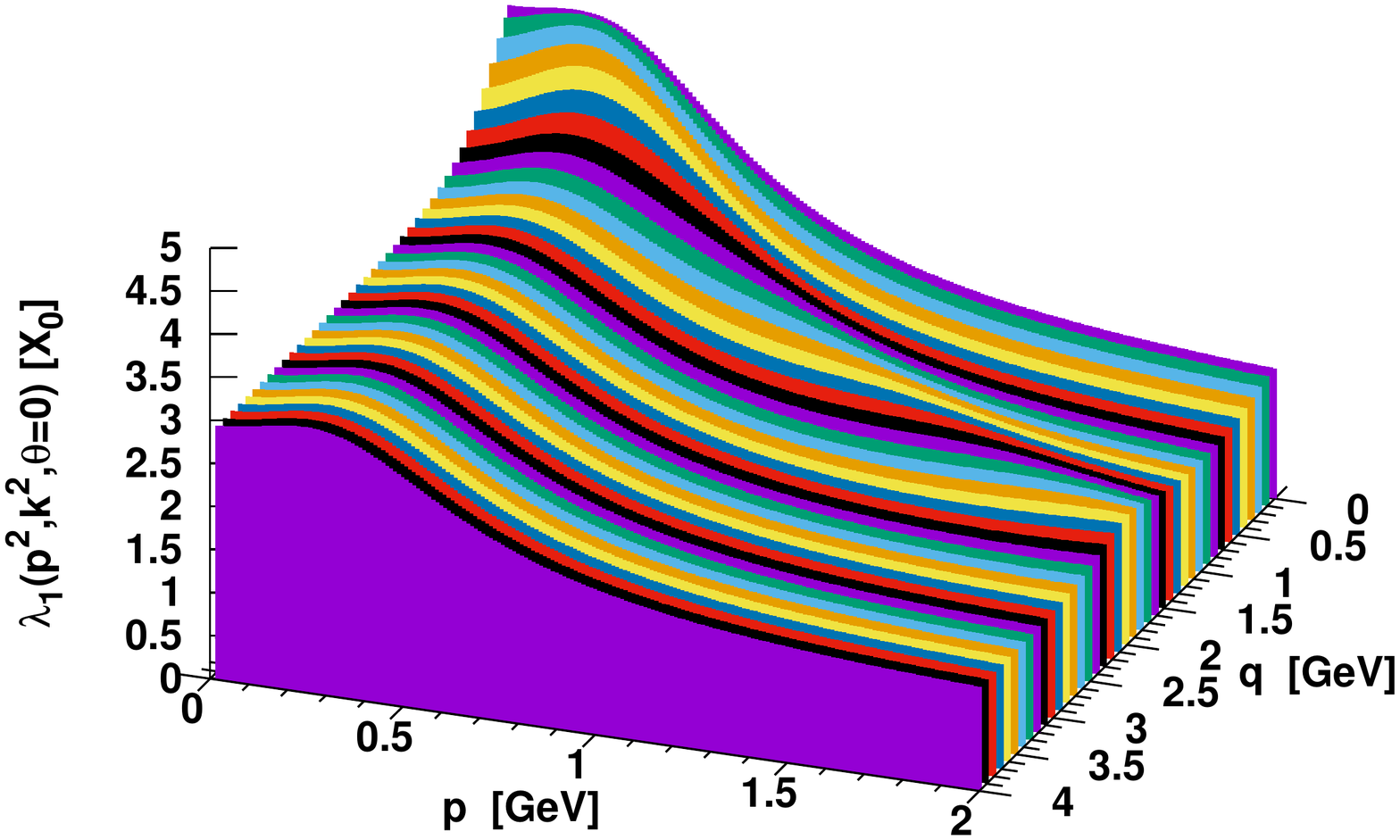}} ~
\subfigure{\includegraphics[width=2.0in]{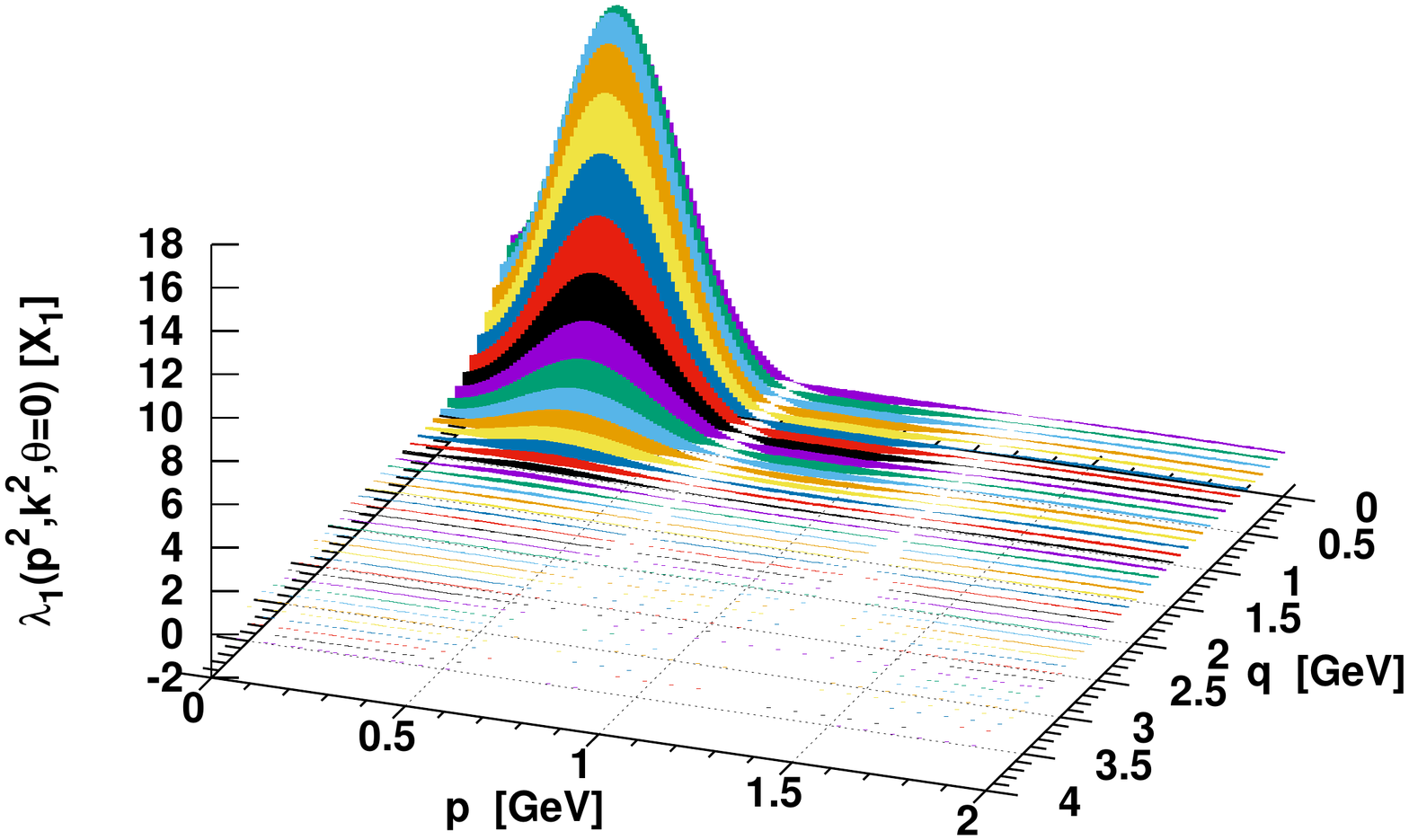}}   
\caption{The contribution proportional to $X_0$ (left) and to $Y_1$ to the form factor $\lambda_1$.} \label{3figs:L1}
\end{figure}

\begin{figure}[h]
\centering 
\subfigure{\includegraphics[width=2.0in]{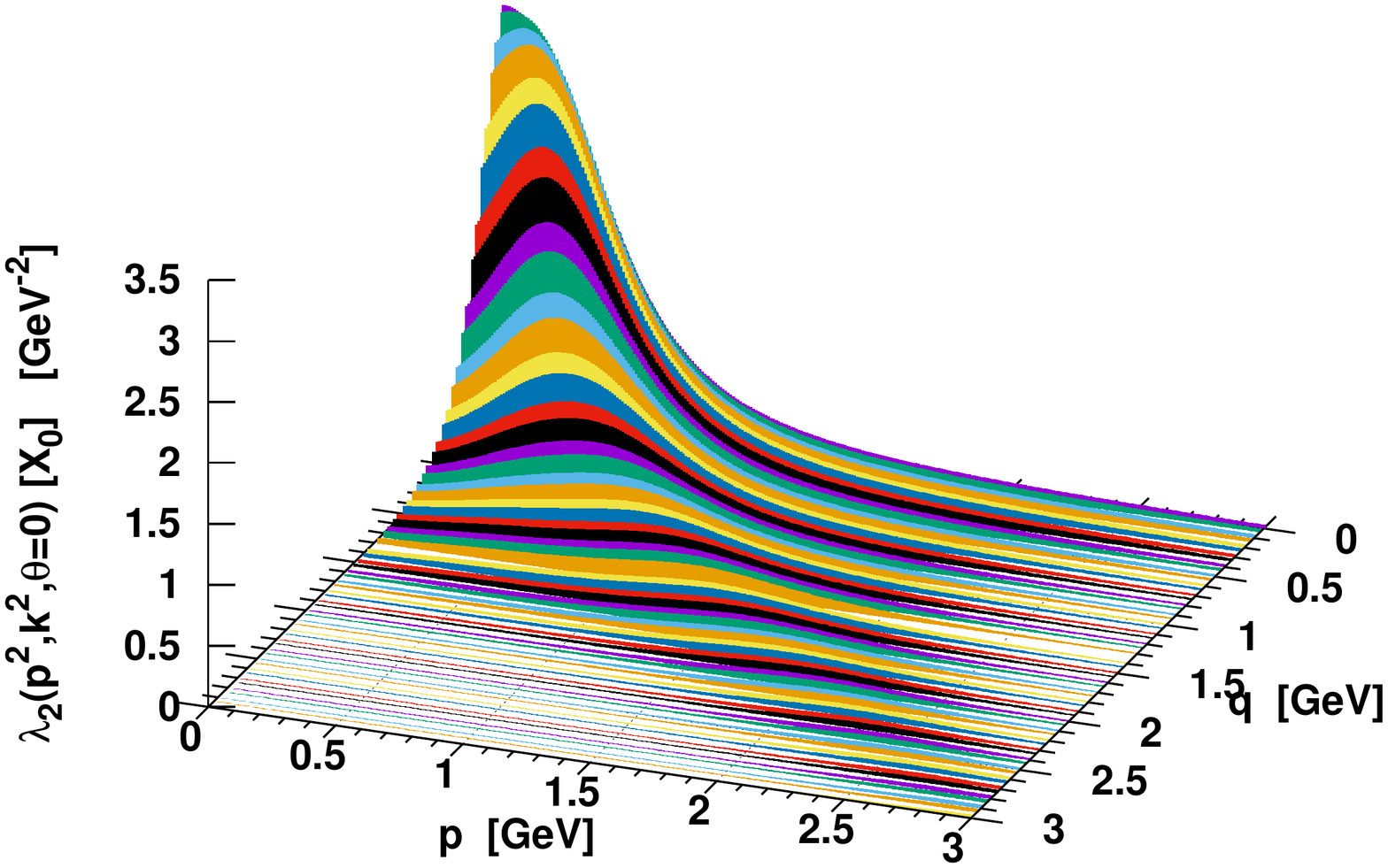}} 
\subfigure{\includegraphics[width=2.0in]{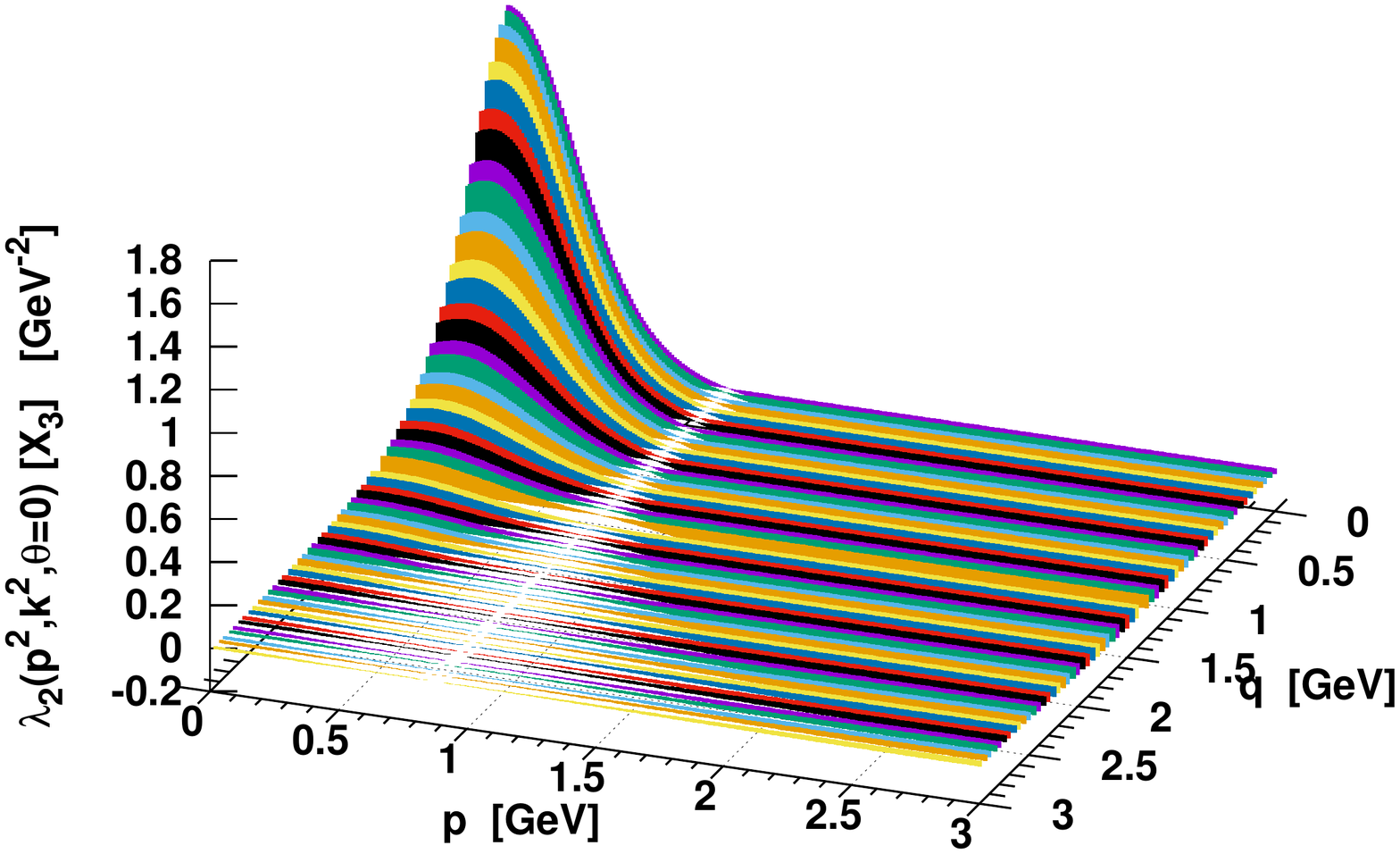}}
\caption{The contribution proportional to $X_0$ (left) and to $Y_3$ to the form factor $\lambda_2$.} \label{3figs:L2}
\end{figure}

\begin{figure}[h]
\centering
\subfigure{\includegraphics[width=2.0in]{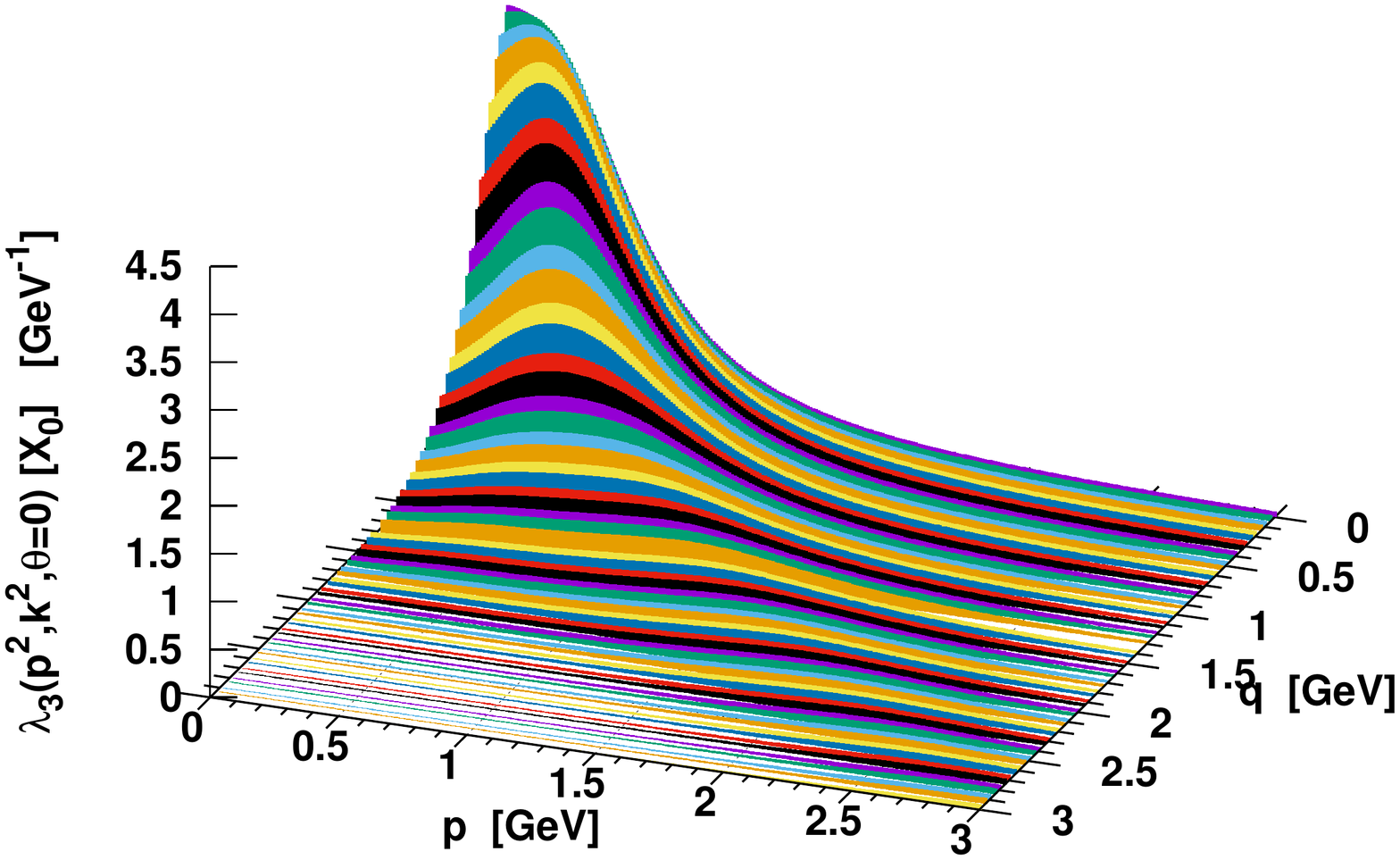}} ~
\subfigure{\includegraphics[width=2.0in]{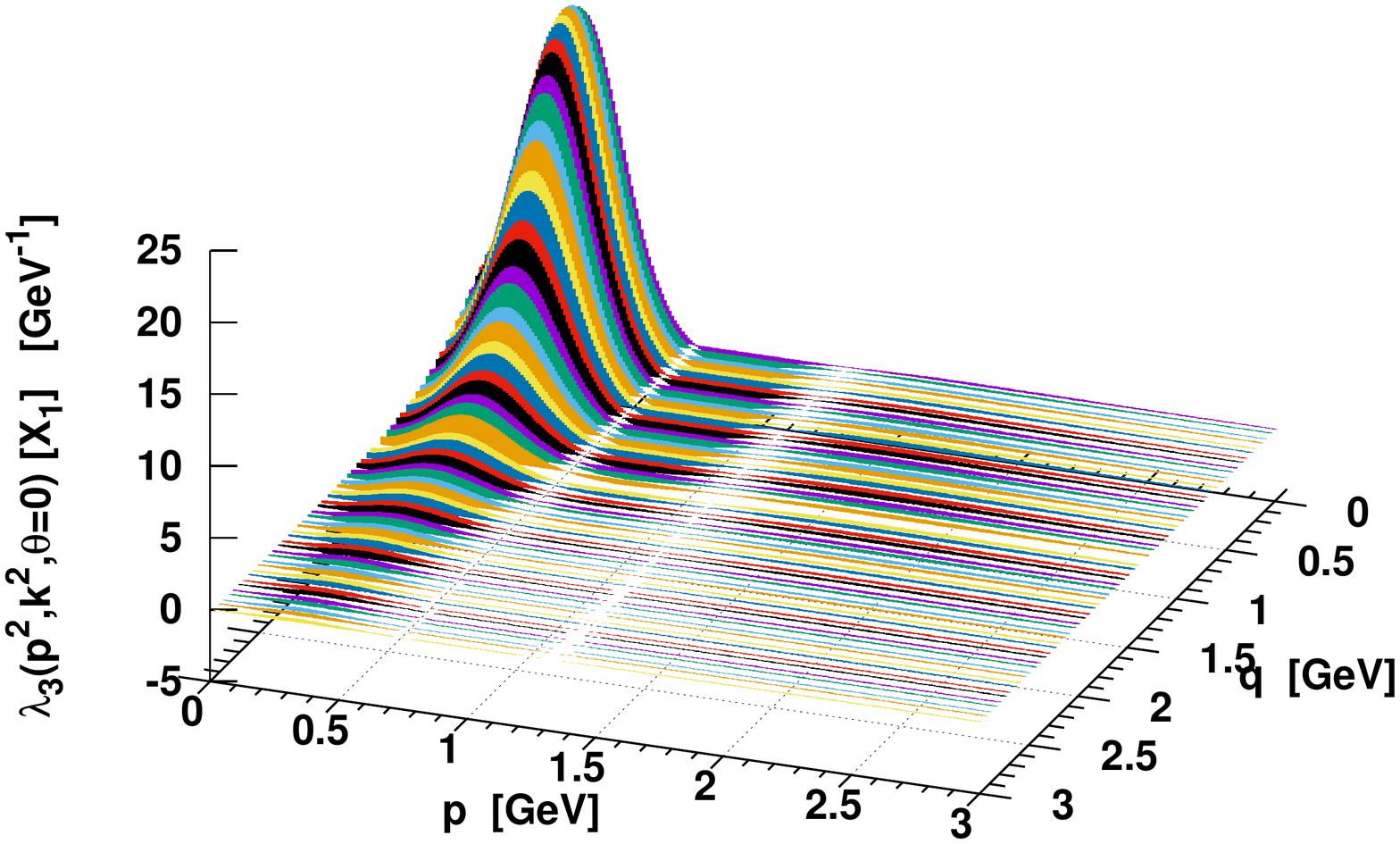}}
\caption{The contribution proportional to $X_0$ (left) and to $Y_1$ to the form factor $\lambda_3$.} \label{3figs:L3}
\end{figure}

\begin{figure}[h]
\centering 
\subfigure{\includegraphics[width=2.0in]{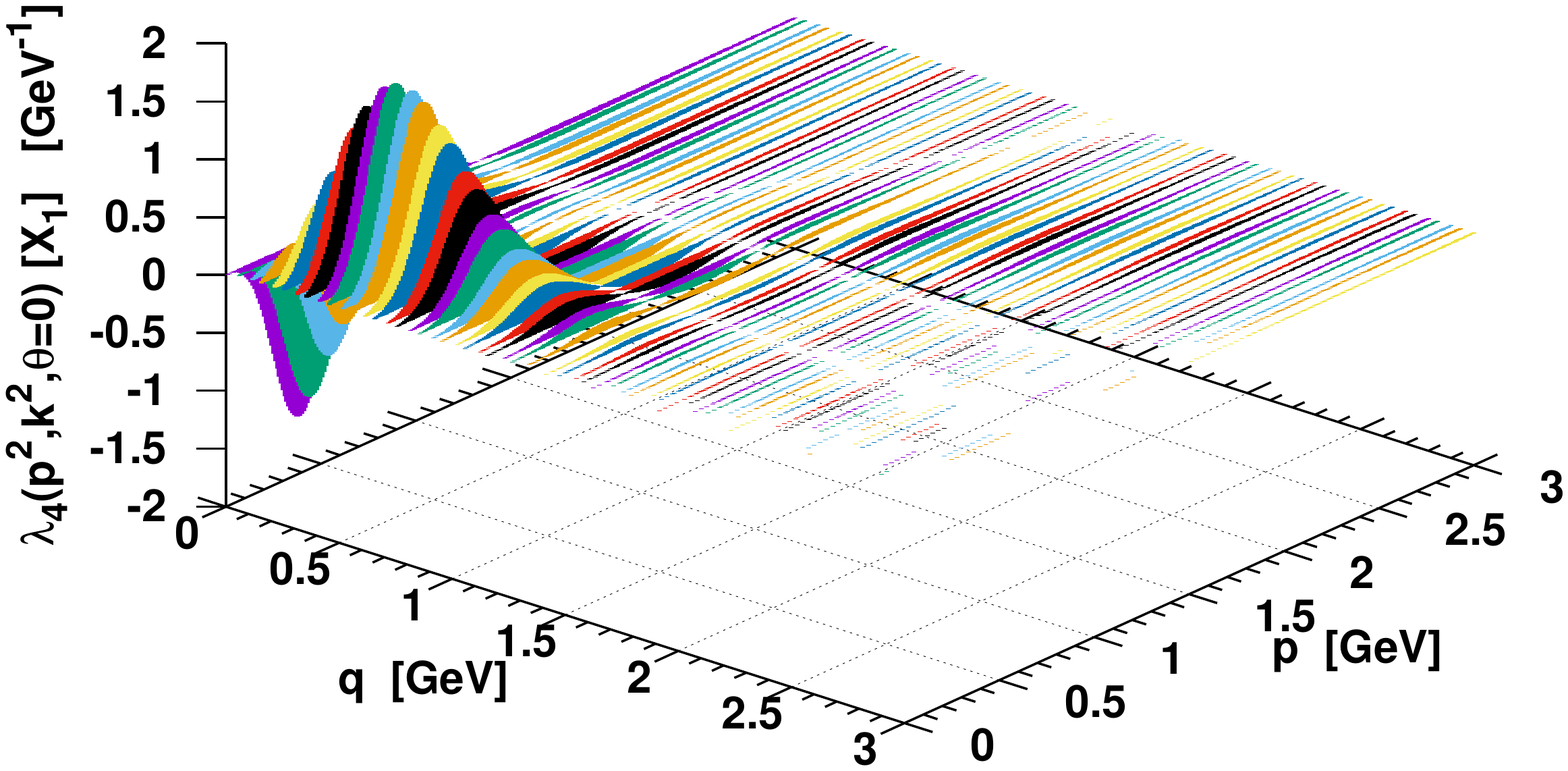}} ~
\subfigure{\includegraphics[width=2.0in]{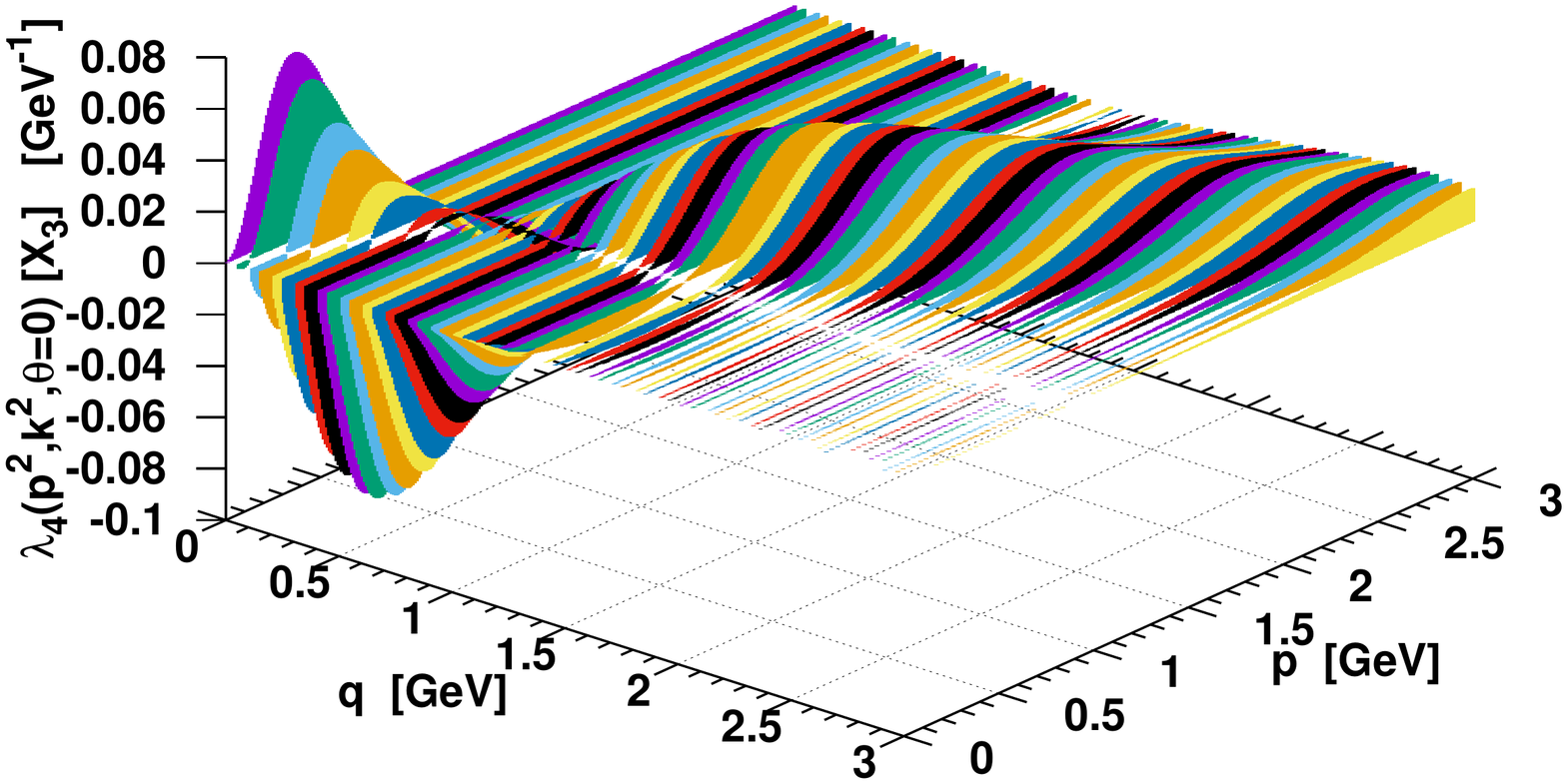}}
\caption{The contribution proportional to $Y_1$ (left) and to $Y_3$ to the form factor $\lambda_1$.} \label{3figs:L4}
\end{figure}

To close our study, in Figs. \ref{3figs:L1} to \ref{3figs:L4} we illustrate the relative importance of the various quark-ghost kernel
form factors to the quark-gluon $\lambda_i$ reported in Fig. \ref{fig:example0} (right). For the dominant form factors $\lambda_1$ and $\lambda_3$ the
major contribution for the infrared enhancement comes from  $X_1$ ($\sim$ 72\% and $\sim$ 83\%), followed by 
$X_0$ ($\sim$ 20\% and $\sim$ 15\%) and a residual contribution from $X_3$ ($\sim$ 8\% and $\sim$ 2\%). The dominance of $X_1$ in
the infrared region for these two form factors was unexpected. $\lambda_2$ is dominated by the $X_0$ contribution ($\sim$ 58\%), followed
by $X_3$ ($\sim$ 30\%) and has a smaller contribution from $X_1$($\sim$ 12\%). Finally, $\lambda_4$ is almost entirely built by the contribution coming
from $X_1$. 
It is somehow unexpected that the various contributions to $\lambda_i$ do not follow the relative importance as suggested by perturbation 
theory. The terms proportional $X_1$ give the major contribution to $\lambda_3$, $\lambda_1$, $\lambda_4$ and $X_0$ is
responsible for the major contribution only to $\lambda_2$.

\section{Discussion and Summary}

The lattice gluon propagator $D(p^2)$ appears as a multiplicative factor in the integral kernels of the Schwinger-Dyson equations
(\ref{Eq:DSEEucScalar}) and (\ref{Eq:DSEEucVector}). 
If $D(p^2)$ changes, the vertex form factors need to change to compensate  the modification of the propagator.
For example, for a term as $D(p^2) \, \lambda_1$, if the propagator $D(p^2)$ is replaced by $D(p^2) f(p^2$), 
then $\lambda_1$ should change accordingly and $\lambda_1 \rightarrow \lambda_1/f(p^2)$.
As there are indications that in full QCD $D(p^2)$ is IR suppressed \cite{Ayala:2012pb,Cui:2019dwv}, meaning that $f(p^2)$ should be smaller than 1 in the infrared region, then the full QCD form factor $\lambda_1$ should increase by a factor $1/f(p^2)$ to keep the solution of the SDE  for the self-energies unchanged.
On the other hand, the gluon propagator also appears in the ansatz\"e for the quark-ghost kernel functions
(\ref{Eq:AnzX0}) - (\ref{Eq:AnzX3}). This functional form is tied to the soft gluon limit of the quark-gluon vertex and the products
appearing in (\ref{Eq:AnzX1}) and (\ref{Eq:AnzX3}) are fixed by the soft gluon limit. Therefore, one expects no or little modifications on the
ansatz for the quark-ghost kernel. This seems to be consistent with the perturbative analysis of the quark-ghost kernel as the inclusion of the
quark contributions are associated with higher order diagrams and are expected to be small. Hopefully this reasoning can be extended to lower momenta.
This analysis of the contribution of the gluon propagator to the general solution of the SDE 
suggests that the form factors given in Fig. \ref{fig:example0} underestimate the $\lambda_i$'s.

Our investigation of the solution of the Schwinger-Dyson equation takes into consideration only the longitudinal part of
the quark-gluon vertex. Of course, the transverse part of the vertex also contributes to the dynamics of QCD and, in principle,
its contribution to the quark gap equation is expected to be within the level of the errors found in the inversion performed herein. 
The transverse form factors have been discussed in the literature, see 
\cite{Aslam:2015nia,Curtis:1990zs,Curtis:1993py,KizPRD2009,Chang:2010hb,Bashir:2011dp,Qin:2013mta,WillEPJA2015,AlbinoPRD2019} and references therein,
and, typically, they are written in a way to take into account the constraints coming
from multiplicative renormalizability and gauge invariance.  

In our description of the longitudinal form factors, see Eqs. (\ref{FFLONG}),  setting $X_0=F(q^2)=1$, $Y_2=Y_3=0$
the vertex reproduces the longitudinal component of the Curtis-Pennington (CP)  fermion-photon vertex~\cite{Curtis:1990zs}.
The CP vertex satisfies the Ward-Green-Takahashi identity and is compatible with multiplicative renormalizability of the  fermion SDE.
The CP vertex also includes a transverse part, associated with the form factor $\tau_6$.  
The CP model was extended later taking into account the photon propagator SDE leading to the Kizilers\"u-Pennington (KP)  
vertex \cite{KizPRD2009}, that requires in addition to the  CP vertex the inclusion of the transverse form factors $\tau_2$, $\tau_3$ and $\tau_8$. 
Our  calculation shows that $\lambda_1$, $\lambda_3$ and $\lambda_4$ are dominated by  $X_1$
against the expectation from the CP and KP vertex models, that suggests a dominance of $X_0$ contribution to the quark-ghost kernel.
 However, our results show, indeed, that the main contribution to $\lambda_2$ comes from $X_0$, in good agreement with the expectations coming from 
the CP and KP vertex models.

The anomalous chromomagnetic form factor, a contribution associated with the transverse form factor $\tau_5$ defined in 
Eqs. (\ref{EQ:Tvertex}) and (\ref{EQ:TFF}), was estimated in \cite{Chang:2010hb}.  
In the infrared region, $\tau_5$ attains a maximum of about 0.20 GeV$ ^{-1}$ and drops one order the magnitude for momentum of $\sim 1$ GeV.  
This form factor can be compared with the longitudinal tensor form factor $\lambda_4$ that,
in the infrared region (below 1 GeV), takes values between -1 and 1.5 GeV$ ^{-1}$.
Of the four longitudinal form factors computed in the current work, the quark-gluon vertex is dominated by the contributions of
$\lambda_1$ and $\lambda_3$,  with $\lambda_4$ being subleading.
However, our estimation of $\lambda_4$ is about 5 times larger than the anomalous chromomagnetic form factor obtained in \cite{Chang:2010hb}. 
If one takes this calculation as an estimation of the typical contribution of the transverse vertex to the quark gap equation, 
one expects small corrections that are due to the transverse form factors that, certainly, will decrease the longitudinal form factors obtained by inversion of the SDE.

Another interesting comparison can be done with the results for the quark-gluon vertex obtained in \cite{WillEPJA2015}, were it is presented a
self-consistent solution of the  SDE for the quark propagator and the quark-gluon vertex function equation in a truncated form, that
include only the three-gluon vertex and the dressed gluon in the decoupling scenario \cite{EichPRD2014}. 
Although in their approach only the dependence on the gluon momentum is kept, their $\lambda_1$ form factor is comparable in magnitude to ours.
In the infrared region,  it attains the maximum value of about 3.5 in the chiral limit for the coupling constant having the value of $\alpha(\mu)=0.7427$ 
\cite{WillEPJA2015}.
The product of these two quantities is $\sim 2.6$ and can be compared to $\alpha_s \, \lambda_1 \, \sim 5$ at the peak of Fig. \ref{fig:example0}
with the two numbers being in the same ballpark. Recall that in our case the vertex depends on the gluon momentum, as in \cite{WillEPJA2015},
but also on the quark momentum and on the angle between the quark and gluon momentum and, therefore, the comparison should be taken with care.
If for our results one averages over the quark momentum, the estimation $\alpha_s \, \lambda_1$ will drop and   the two results  become closer.
The differences in $Z(p^2)$ at the small-$p^2$ between our quark wave function, see Fig.  \ref{fig:DSE_rhs}, and that computed in \cite{WillEPJA2015}
are overcome by the large vertex enhancements. Moreover, the lattice gluon dressing function is magnified in the IR with respect to the decoupling scenario used in the later work. 
 
Our  description of the quark wave function and of the the running mass reported in Fig.~\ref{fig:DSE_rhs} should also be compared
to the earlier studies of the quark propagator within the rainbow-ladder approximation. 
In \cite{FischerPLB2002,FischerPRD2003} a coupled set of renormalized Landau gauge truncated DSE's for the quark, gluon, and ghost propagators
was solved using a bare quark-gluon vertex, the Ball-Chiu and Curtis-Pennington models with and without nonabelian correction for the vertex.
The framework used has a number of parameters that can be varied and it is possible to explore the parametric freedom of the models
to obtain a  running  quark mass and a quark wave function that is similar to what is seen in Fig.~\ref{fig:DSE_rhs}. This
is linked with a strong infrared enhancement of the kernel of the SDE.   
Our approach uses a clear grow of the gluon and ghost dressing functions, taken from recent lattice simulations, that via the Slavnov-Taylor identity
and the modeling of the longitudinal  form factors are fundamental to give the right strength of the SDE kernel in the infrared region.

The rainbow-ladder approach of  Maris and Roberts \cite{MarisPRC1997}, further developed by
Maris and Tandy \cite{MarisPRC1999} proposes a quark-quark scattering kernel ansatz\"e, that 
has a strong IR enhancement to break dynamically the chiral symmetry,
in a framework where the the axial-vector Ward-Takahashi identity and the Gell-Mann-Oakes-Renner relation  are exactly satisfied.
In addition the model preserves the one-loop renormalization group structure of QCD. 
This approach was explored in~\cite{BhagwatPRC2003} and a fair description of the Lattice results for the running quark  mass 
was achieved. Interesting enough, the study concluded that the dressed-gluon is not enough to obtain the existing lattice quark dressing data, 
unless the SDE kernel is enhanced for infrared momenta through  the dressing of the quark-gluon vertex. 
Furthermore, the vertex in these works would correspond to the contribution of $\lambda_1$, which we found to have a quite
strong IR enhancement as seen in Fig.~\ref{fig:example0}; recall that in our case the form factors are function not only on the gluon momentum,  
but also on the quark momentum.
The symmetry-preserving Maris-Tandy model has been very successful in describing meson and baryon properties 
(see e.g. the reviews \cite{CloetPPNP2014,EichmannPPNP2016}), emphasizing 
the role of the IR enhancement of the quark-gluon vertex as necessary  to reproduce the hadron phenomenology.

In summary, the inclusion of the soft-gluon limit information in our analysis enhances the infrared region
of the longitudinal form factors of the quark-gluon vertex.
Furthermore, the relative contribution of the quark-ghost kernel form factors to the quark-gluon vertex do not follow in the infrared region  
their relative importance found in the ultraviolet region. 
Our solution for the longitudinal vertex form factors  given in Eq.~(\ref{FFLONG}) together with the parametrization (\ref{PADE_X3})
and the results reported in Table  \ref{tab:padefit}, describes analytically all the components of the vertex considered, allowing easily further applications.
It still remains a challenge to include the transverse components of the quark-gluon vertex. 
The error observed in the SDE, at the level of 4\%, is expected to be associated 
not only with the limitation of the ansatz but also with the missing transverse form factors.
The  phenomenological implications of the present findings to the meson structure is unexplored and will be considered in a future publication.

{\it Acknowledgements.}
The Coimbra Physics Centre is supported by FCT through the project UID/FIS/04564/2016, also co-funded by COMPETE-UE.
This work was supported in part by the project INCT-FNA Proc. No. 464898 /2014-5;  by CAPES - Finance Code 001 (CAPES-PRINT 88887.374225 /2019-00) and under grant 88881.309870/2018-01 (WP); by Conselho Nacional de Desenvolvimento Cien\-t\'\i \-fico e Tecnol\'ogico (CNPq) under grants No. 308486/2015-3 (T.F.), 438562/2018-6 and 313236/2018-6 (WP) and;
by Funda\c c\~ao de Amparo \`a Pesquisa do Estado de S\~ao Paulo (FAPESP) under
the thematic Projects No. 2013/26258-4 and No. 2017/05660-0.

\begin{appendix}
\section{Lattice ghost, quark and gluon propagators \label{Sec:propagators}}

For completeness, here we report on the fitting expressions used to model the lattice data for the various propagators. Note that, in all cases the model function incorporates
the perturbative behaviour at high momenta. Details about how the model functions were computed and how well they reproduce the lattice propagator data can
be found in \cite{Oliveira:2018ukh}.

The lattice Landau gauge gluon propagator and global fits to the data was published in \cite{Dudal:2018cli}. In our analysis we use
the following model function for the gluon propagador (see \cite{Dudal:2018cli} for further details)
\begin{equation}
   D (p^2)  =  Z \, \frac{ p^2 + M^2_1}{p^4+ M^2_2 \, p^2 + M^4_3} \left[ \omega \,  \ln \left(\frac{p^2 + m^2_0}{\Lambda^2_{QCD}}\right) + 1 \right]^{\, \gamma} \ ,
   \label{Eq:global_gluon_Fit}
\end{equation}
where the gluon anomalous dimension is
$\gamma = -13/22$, $Z  = 1.36486 \pm 0.00097$, $M^2_1   =  2.510 \pm 0.030 \mbox{ GeV}^2$, $M^2_2   =  0.471 \pm 0.014 \mbox{ GeV}^2$, 
$M^4_3   =  0.3621 \pm 0.0038  \mbox{ GeV}^4$, $m^2_0   =  0.216 \pm 0.026  \mbox{ GeV}^2$. In Eq. (\ref{Eq:global_gluon_Fit})
$\omega = 33 \,  \alpha_s ( \mu ) / 12 \pi$  and to reproduce the lattice data $\Lambda_{QCD} = 0.425 \mbox{ GeV}$ and the strong coupling constant 
reads $\alpha_s ( \mu = 3 \mbox{ GeV} ) = 0.3837$. We call the reader attention that in the inversion of Dyson-Schwinger equation the propagator
is renormalised, in the MOM scheme, at a difference scale. This implies the computation of a global renormalisation constant that multiplies the above
$D(p^2)$. The same  reasoning applies to the ghost propagator and to the quark wave function.

For the ghost propagator we take the data reported in~\cite{Duarte:2016iko} for the simulation using a $80^4$ and use the following functional
form to describe the lattice data
\begin{eqnarray}
   & & D_{gh} (p^2)  =  \frac{F(p^2)}{p^2} \nonumber \\
   & & \qquad =  \frac{Z}{p^2} ~ \frac{ p^4 + M^2_2 \, p^2 + M^4_1}{p^4+ M^2_4 \, p^2 + M^4_3}  \times \nonumber \\
   &  & \qquad\qquad\quad \times \, 
            \left[ \omega \,  \ln \left(\frac{p^2 + \frac{m^4_1}{p^2 + m^2_0}}{\Lambda^2_{QCD}}\right) + 1 \right]^{\, \gamma_{gh}} .
   \label{Eq:global_ghost_Fit}
\end{eqnarray}
The fit to the lattice data results in $Z  =  1.0429 \pm 0.0054$, $M^4_1 = 18.2 \pm 5.7 \mbox{ GeV}^4$, $M^2_2 = 33.4 \pm 6.4 \mbox{ GeV}^2$,
$M^4_3 = 6.0 \pm 2.7 \mbox{ GeV}^4$, $M^2_4 = 29.5 \pm 5.7$ GeV$^2$,
$m^4_1 = 0.237 \pm 0.049$, $m^2_0 = 0.09 \pm 0.42 \mbox{ GeV}^2$.
The ghost anomalous dimension reads $\gamma_{gh} = - 9/44$ and
$\omega$ and $\Lambda_{QCD}$ take the same values as in the gluon fitting function (\ref{Eq:global_gluon_Fit}).

For the quark propagator we consider the result of a $N_f = 2$ full QCD simulation in the Landau gauge~\cite{Oliveira:2016muq,Oliveira:2018lln} for
$\beta = 5.29$, $\kappa = 0.13632$ and for a $32^3 \times 64$ lattice. For this particular lattice setup,
the corresponding bare quark mass is 8 MeV and the pion mass reads $M_\pi = 295$ MeV. 
For the quark wave function $Z(p^2)$ we corrected the lattice data to become compatible with perturbation theory at high momenta, see
 \cite{Oliveira:2018ukh} for details, and use the following functional form
\begin{equation}
    Z(p^2) = Z_0 ~ \frac{ p^4 + M^2_2 \, p^2 + M^4_1 }{p^4 + M^2_4 \, p^2 + M^4_3} \label{quark_Z_function}
\end{equation} 
with $Z_0 = 1.11824 \pm 0.00036$, $M^4_1 = 1.41 \pm 0.18 \mbox{ GeV}^4$, 
$M^2_2 = 6.28 \pm1.00 \mbox{ GeV}^2$, $M^4_3 = 2.11 \pm 0.28 \mbox{ GeV}^4$,
$M^2_4 = 6.20$ $\pm 0.98$ GeV$^2$. 
The choice of the functional form for the running quark mass was
\begin{equation}
 M(p^2) = \frac{ m_q ( p^2) }{ \left[ A + \log( p^2 + \lambda \, m^2_q(p^2) ) \right]^{\gamma_m} }
 \label{Eq:RunningMass}
\end{equation}
where $\gamma_m = 12/29$ is the quark anomalous dimension for $N_f = 2$ and
\begin{equation}
        m_q(p^2) = M_q ~\frac{ p^2 + m^2_1}{ p^4 + m^2_2 \, p^2 + m^4_3 } + m_0 \ .
         \label{Eq:RunningQuarkMass}
\end{equation}
The parameters in Eqs. (\ref{Eq:RunningMass}) and (\ref{Eq:RunningQuarkMass}) are 
 $M_q = 349 \pm 10 \mbox{ MeV GeV}^2$, $m^2_1  = 1.09 \pm 0.43 \mbox{ GeV}^2$, $m^2_2  = 0.92 \pm 0.28 \mbox{ GeV}^2$, 
$m^4_3 = 0.42 \pm 0.15 \mbox{ GeV}^4$,
$m_0 = 10.34 \pm 0.63$ MeV and $A = -2.98 \pm 0.25$ 
and $\lambda = 1 \mbox{ GeV}^2/\mbox{MeV}^2$.

\end{appendix}


\end{document}